\documentclass[11pt,twoside]{article}
\usepackage[]{graphicx}
\usepackage{natbib}
\usepackage{fancyhdr,setspace,subfigure,rotate}
\usepackage{amsthm,amssymb,amsmath}
\usepackage{amsmath}
\usepackage{lineno}
\usepackage{color}
\usepackage{ctable}
\IfFileExists{url.sty}{\usepackage{url}}{\newcommand{\url}{\text}}
\usepackage{sectsty}
\allsectionsfont{\normalsize}
\numberwithin{equation}{section}
\usepackage{caption}
\usepackage{float}
\usepackage[ruled,vlined]{algorithm2e}
\usepackage{algorithmic}
\usepackage{outlines}
\usepackage{dblfloatfix}

\begin{document}
\title{\large\bfseries Optimal Sampling Regimes for Estimating Population Dynamics}
\author{
 Rebecca E.~Atanga$\!^1$, Edward L.~Boone$\!^1$, Ryad A.~Ghanam$^2$, Ben Stewart-Koster$\!^3$ \\[0.2in] 
\normalsize\itshape $^1$Department of Statistical Sciences and Operations Research,\\ 
\normalsize\itshape Virginia Commonwealth University, Richmond, VA 23284, USA\\[0.05in]
\normalsize\itshape $^2$Virginia Commonwealth University - Qatar,\\ 
\normalsize\itshape Education City, Doha, Qatar.\\
\normalsize\itshape $^3$Australian Rivers Institute,\\ 
\normalsize\itshape Griffith University, Southport, QLD, Australia.\\
}

\date{\today}
\maketitle

\hrule
\begin{abstract}

Ecologists are interested in modeling the population growth of species in various ecosystems. Studying population dynamics can assist environmental managers in making better decisions for the environment. Traditionally, the sampling of species and tracking of populations have been recorded on a regular time frequency. However, sampling can be an expensive process due to available resources, money and time. Limiting sampling makes it challenging to properly track the growth of a population. Thus, we propose a new and novel approach to designing sampling regimes based on the dynamics associated with population growth models. This design study minimizes the amount of time ecologists spend in the field, while maximizing the information provided by the data.

\end{abstract}
\vspace{0.1in}
\hrule

\section{Introduction}
\label{sect-intro}

Population growth models are commonly used in ecology when studying plants, animals, and organisms as they grow over time and interact with the environment. In earlier research, \citet{Hsu1984} evaluate seedling growth under laboratory and field conditions. According to \citet{Huang2016}, early life-cycle events play critical roles in determining the dynamics of plant populations and their interactions with the surrounding environment. Germination patterns modeled by growth help ecologists understand the diversity, variation, and climate change within a system. \citet{Gamito1998} further emphasize the importance of modeling growth and community dynamics using applications to fish populations. Population dynamics of fish are commonly studied to learn about water quality and the environment.

In riverine settings, the health of a waterway can be determined by the abundance of fish present. \citet{Kennard2005} study species that are highly tolerant of human-induced disturbances and consider certain fish as good indicators of river health. Ecologists model the population growth of indicator species to preserve habitats as land is developed along the banks of rivers. Not only can population growth models help preserve the environment, but also ecologists model growth of crops to improve economies. In Vietnam, rice-shrimp farming highly impacts the economic security of the country as these are the two main crops in the region. \citet{Leigh2017} study the environmental conditions of ponds used for harvesting and address the risks that affect the survival and yields of crops. In order to improve year-round farming, ecologists monitor the growth and abundance of shrimp and rice. 

Whether researchers are interested in improving yields for farmers, preserving the health of rivers or studying germination of seedlings, collecting ecological data has always been a time consuming and expensive process. Ecologists fishing in rivers face strenuous terrain, varying water levels, and dangerous wildlife and prefer to limit their time collecting samples. Though farmers raise crops in controlled environments, limiting sampling costs can increase revenue. During germination studies, scientists prefer to optimize their time and resources required to grow samples. In all cases, researchers prefer to reduce costs when collecting data by decreasing their necessary sample size without compromising the ability to accurately track growth. 

Rather than traditional sampling methods, we propose an approach that designs optimal sampling regimes based on the dynamics associated with a population. Though traditional methods are often convenient, statistically designing an experiment can optimize sampling procedures for ecologists without compromising the information obtained from the data. This paper investigates capturing the dynamics of ecological models by designing optimal sampling regimes. Given our approach is intended to design experiments prior to data collection, we simulate realistic growth using a well-known ecological model with set parameters that can later be replicated with minimal error. 

Using Bayesian inference, we can predict the probability of the model parameters based on simulated data by employing Markov Chain Monte Carlo (MCMC) sampling techniques demonstrated by \citet{Gilks1995}. We plan to learn about the system in a sequential manner by using various criteria to optimize the system. The Bayesian model combined with design of experiment techniques can be used to determine the optimal sampling frequency that minimizes costs associated with data collection while obtaining the maximum amount of information from the data. The statistical methods in this paper are applied to models with different growth rates using various criteria to demonstrate the versatility of our approach. As a design study, our goal is to develop a methodology that provides ecologists with optimal times to collect data that accurately capturing the population growth dynamics of the system.  

\section{Methods}
\subsection{Statistical Model}
The well-known logistic equation was first developed by \citet{Verhulst1838}. The rediscovery of the equation by \citet{Reed1927} is often used in ecology to represent the population dynamic where the rate of reproduction is proportional to both the existing population and the amount of available resources.  This model is popular in ecology due to the realistic self-limiting growth of a population and the existence of its analytic solution. The logistic equation is written in the mathematical form

 \begin{equation}
  \frac{dN}{dt} = rN \left( 1 - \frac{N}{K} \right)
    \label{eq:1}
\end{equation}
with solution
 \begin{equation*}
  N(t) = \frac{KN_0}{(K-N_0)e^{-rt} +N_0}.
\end{equation*}

\noindent The variables $N$ and $t$ represent population size and time respectively. Parameter $r$ is the population growth rate, and parameter $K$ is the maximum level that the population can reach also known as the carrying capacity. $N_{0}$ represents the population at time $t=0$ referred to as the initial population. Given that ecologists have expertise in modeling logistic growth, the Bayesian approach can be used to incorporate prior knowledge of the system. Thus, the ecological model is specified probabilistically in the Bayesian framework. 

Given our well-defined model, we implement the Bayesian approach using Bayes' Theorem \citep{Bayes1763}. The theorem combines prior beliefs with the likelihood of an experiment to determine a posterior belief. The prior belief of a parameter $\theta$ is noted by $\pi(\theta)$. The likelihood of the experiment is a conditional probability of the data $x$ given the parameter $\theta$ noted by $L(x| \theta)$. The posterior belief is given by the continuous case of Bayes' Theorem, also known as the posterior distribution. The posterior distribution defines the conditional probability of a parameter $\theta$ given the observance of data $x$, which is written mathematically as

\begin{equation}
  P(\theta | x) = \frac{L(x|\theta) \pi(\theta)}{\int_{\Theta}L(x|\theta) \pi(\theta)d\theta} .
\end{equation}

\noindent Specifying equation (\ref{eq:1}) as a Bayesian model requires the specification of the likelihood of the experiment and the prior distributions for each parameter. The nature of the logistic equation tracks the population of a species $N$ across a fixed time $t$ where the growth rate parameter $r$ and carrying capacity $K$ are independent fixed values. The population at a specified time $N_i|t_i$ is assumed to have a Poisson likelihood given that the observations $N_i$ are independent counts occurring at known times $t_i$. The analytic solution to the logistic equation (\ref{eq:1}) provides the expected value of the likelihood set as $\lambda_i = N(t_i) $. This provides the conditional likelihood of the experiment 
\begin{equation}
N_i|t_i  \sim Poisson( \lambda_i | t_i ),
\label{like}
\end{equation}

\noindent where $\lambda_i$ depends on the parameters $r$ and $K$. Thus, the prior distributions must be specified for the growth rate and carrying capacity parameters. 

When modeling logistic growth, ecologists are aware that a population can only reach a maximum capacity by increasing at a positive rate. This implies that the growth rate $r$ and carrying capacity $K$ must be positive values. Though many distributions have positive supports, the log-normal distribution easily specifies numerical information by using the mean and variance. Less informative priors such as the gamma distribution can cause model instability, which defeats the purpose of incorporating prior beliefs.  Thus, informative priors are specified as
\begin{equation}
\begin{split}
r &\sim lnnorm(1, 10) \\
K &\sim lnnorm(2000, \frac{1}{10}), \\
\end{split}
\label{eq:prior}
\end{equation}

\noindent where the carrying capacity is assumed to reach a maximum of two thousand, while the expected growth rate is centered about one. These explicit priors are implemented to represent expert knowledge of the system and will be used for all simulations in this paper. The Bayesian approach emphasizes combining informative prior beliefs with experimental knowledge to determine a posterior belief. Thus, the specified model will be used to predict the parameters of various realistic population growth models that can be optimized by assessing various criteria.

\subsection{Optimal Designs}

Optimal designs are used to estimate statistical models while reducing experimental costs. The objective of optimal design is to eliminate uncertainty by minimizing the variability of the parameter estimates. Traditionally, optimal designs \citep{Kiefer1959} minimize the variance of the parameter estimates while maximizing the information matrix. Specifically, the Fisher information matrix is defined as the negative expectation of the second derivative of the log-likelihood function with respect to the parameter $\theta$ 
 \begin{equation}
 F(\theta) = -E[ \frac{\partial^2}{\partial \theta^2} logL(x|\theta)],
 \label{eq:f}
 \end{equation}

\noindent where the expectation is taken over the sample space of the observations $x$ and parameter space of $\theta$. Optimality criteria are applied to the Fisher information matrix $F(\theta)$ and provide measures of fit to assist with model selection. The popular $A$ and $D$ optimal designs are commonly used in experimentation due to their ability to limit computational expenses. $A$ optimality criterion minimizes the trace of the inverse of the Fisher information matrix, which equivalently minimizes the average variance of the parameter estimates of a model. Whereas, $D$ optimal designs consist of minimizing the determinant of the inverse of the Fisher information matrix, again minimizing the parameter estimates of the model. 
 \begin{equation}
  \begin{tabular}{l}
 $A$ - Optimal: $ \min tr (F^{-1}(\theta)) $ \\
 $D $ - Optimal: $ \min | F^{-1}(\theta) |$ \\
  \end{tabular}
 \end{equation}

Though these designs are commonly used in practice, $F(\theta)$ can be difficult to calculate when the model parameters are unknown. Instead of estimating the Fisher information matrix using parameter estimates noted by $\hat\theta$, we consider the variance of the parameter estimates. The inverse of the estimated Fisher information matrix $F(\hat\theta)$ is an estimator of the asymptotic covariance matrix \citep{abt1998}.
 \begin{equation}
 \Phi = Var(\hat\theta) = [F(\hat\theta)]^{-1}
 \end{equation}
 
 \noindent The covariance matrix is much easier to calculate than the Fisher information matrix for certain models. Thus, defining $F(\hat\theta)$ in terms of $\Phi$ redefines the optimality criteria in terms of the estimated covariance matrix. I-optimal designs \citep{Atkinson2007} known for minimizing the average prediction variance over the entire design region are also considered. 
 \begin{equation}
  \begin{tabular}{l}
 $A_{\Phi}$ - Optimal: $ \min tr ( \Phi ) $ \\
 $D_{\Phi}$ - Optimal: $ \min | \Phi |$ \\
 I - Optimal : $\min \bar \Phi_{pred}$
  \end{tabular}
   \label{eq:opt}
 \end{equation}

$\bar \Phi_{pred}$ notes the average prediction variance of the model. Rather than minimizing the trace of the inverse of the Fisher information matrix, $A_{\Phi}$-optimal designs minimize the trace of the covariance matrix. $D_{\Phi}$-optimal designs minimize the determinant of the covariance matrix. While, I-optimal designs minimize the average prediction variance over the design space. Again, optimality criteria are measures of fit used to guide optimization processes. $A_{\Phi}$, $D_{\Phi}$ and I optimality criteria are used in the proposed sequential optimality algorithm presented in the next section.

\subsection{Sequential Optimality}

In practice, design techniques are used prior to collecting data in order to optimize the amount of information obtained during experimentation. Thus, we propose a new and novel approach to designing experiments that predicts the next best design point based on current information of the system. Our algorithm is an adaptation of the well-known simulated annealing algorithm applied by \citet{Laarhoven1987}. Simulated annealing searches a discrete design space for the global optimum of the system. Our proposed method sequentially explores subsets of the design space in search of the global optimum within a set window of time.

\begin{center}
\begin{algorithm}[H]
\SetAlgoLined
\textbf{Begin} \\
\hskip \algorithmicindent Choose an initial design $t_1, ..., t_n$\\
\hskip \algorithmicindent Set a design budget, $b$ \\
\hskip \algorithmicindent Set a design window, $w$ \\
\hskip \algorithmicindent Set criteria, $C$: \\
\hskip \algorithmicindent \hskip \algorithmicindent (i.)  $A_{\Phi} = \min tr ( \Phi ) $ \\
\hskip \algorithmicindent \hskip \algorithmicindent (ii.)  $D_{\Phi} = \min | \Phi |$ \\
\hskip \algorithmicindent \hskip \algorithmicindent (iii.)  I = $\min \bar \Phi_{pred}$ \\
\hskip \algorithmicindent \hskip \algorithmicindent \hskip \algorithmicindent \textbf{For} $ D = t_1, . . . ,t_n$: \\
\hskip \algorithmicindent \hskip \algorithmicindent \hskip \algorithmicindent \hskip \algorithmicindent (a) Draw a sample $t^* = \{ t_{n+1}, ..., t_{n+w} \}$\\
\hskip \algorithmicindent \hskip \algorithmicindent \hskip \algorithmicindent \hskip \algorithmicindent (b) Accept the new state $t_{new}= \min \arg (t^*_c)$\\
\hskip \algorithmicindent \hskip \algorithmicindent \hskip \algorithmicindent \textbf{Repeat} until $D = t_1, ..., t_b$\\
\textbf{End}
\caption{Sequential Optimality}
\label{1}
\end{algorithm} 
\end{center}

The algorithm begins by setting an initial design of size $n$ typically set as the number of parameters in the system plus one. Then, a design point budget, $b$, is chosen according to the number of runs the experimenter can afford. Once the initial data is collected and a design budget is set, the design window, $w$, needs to be determined. The design window can be established by dividing the planned time interval, $T$, by the design point budget. For example, this study focuses on sampling across a one hundred day season with a design budget of size ten. Thus, a practical design window would be to search ten days into the future. The window of points following the current design $D$ are explored as candidate samples. Each candidate is evaluated by running the Metropolis-Hastings (MH) algorithm \citep{Metropolis1953b} guided by specified optimality criterion, $C$. The optimal point in the window $t_{new}$ is added to the design, and the process repeats until the design point budget, $b$, is exhausted.
 
The sequential optimality algorithm is intended to search the design space of temporal models. The specified Bayesian model in Section 2.1 tracks the population of a species across one hundred time points with a carrying capacity of two thousand. The simulation study involves different growth rate parameters and simulates data using MCMC sampling to predict the probability of the model parameters. Then, the proposed method is implemented using all design criteria from equation (\ref{eq:opt}) for each model to demonstrate how to predict optimal future design points. The purpose of this design study is to optimize sampling procedures for ecologists by learning about systems in a sequential manner. 

\section{Simulation Study}

This simulation study demonstrates methods that can be used to design sampling schedules for ecologists based on varying population dynamics. Real data cannot be used due to the fact that this experimental study has not yet been implemented in practice. Thus, data is simulated for normal, fast and slow growth models across one hundred time points. The normal growth rate is set to 10\%, the fast rate is set to 100\%, and the slow rate is set to 5\% growth. All models have a maximum carrying capacity of 2000, and ten thousand MCMC samples are used to predict the probability of the model parameters. Based on the planned time interval of one hundred days, all optimal designs will consist of ten points that search windows of size ten. This allows the entire budget to be used should the farthest point in each window be selected. Therefore, ten design points are also selected during simulated annealing for comparison purposes. Simulated annealing is implemented first to demonstrate the global optimum for each scenario as a reference when comparing sequential optimality. Then, sequential optimality is demonstrated as a novel approach to designing optimal sampling regimes. 

All ground truth models are generated using the logistic equation and analytic solution provided by equation (\ref{eq:1}). Each simulation plots the ground truth model as a black curve tracking the population size across time. 10,000 MCMC samples are used for each simulation, and the optimal design points selected are plotted as blue open circles. The designs are fit by a red curve with the prediction interval plotted as red dotted lines calculated by the 2.5\% and 97.5\% quantiles of the predicted values. It will become clear by the results that there are many ways to successfully design experiments for ecological models. 

\subsection{Simulated Annealing}

Simulated annealing is implemented as an adaptation of the Metropolis-Hastings algorithm \citep{Metropolis1953b} to explore the design space for an optimal solution. Considering normal, fast and slow growth rates, the specified optimality criteria from equation (\ref{eq:opt}) are used to guide the algorithm and find global optimums. As stated previously, the models simulate growth at 10\%, 100\%, and 5\% rates. Each criterion guides the simulated annealing process to produce optimal designs for each model.

\begin{figure}[H]
\centering
\subfigure[]{
\includegraphics[width=1.96in]{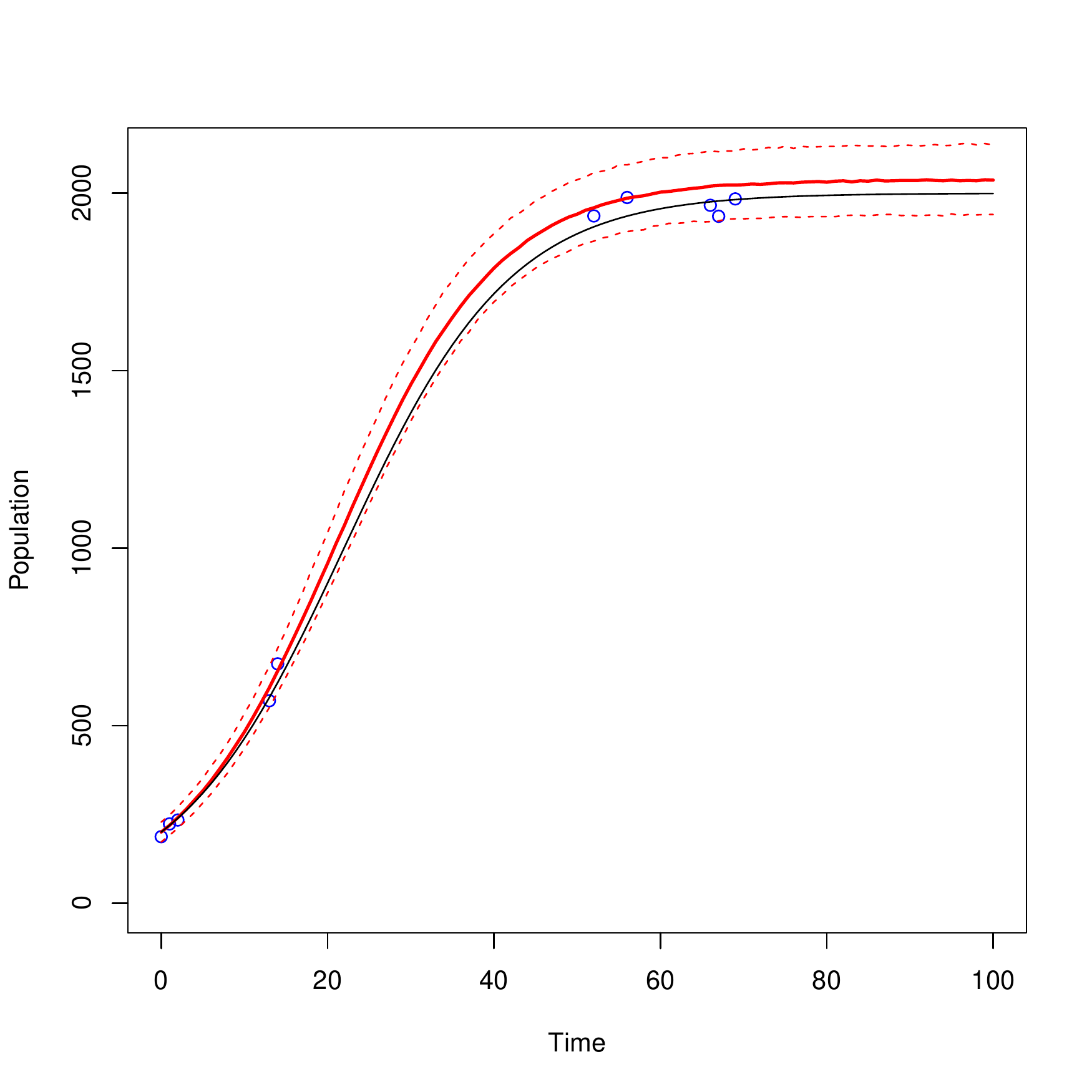}
}
\subfigure[]{
\includegraphics[width=1.96in]{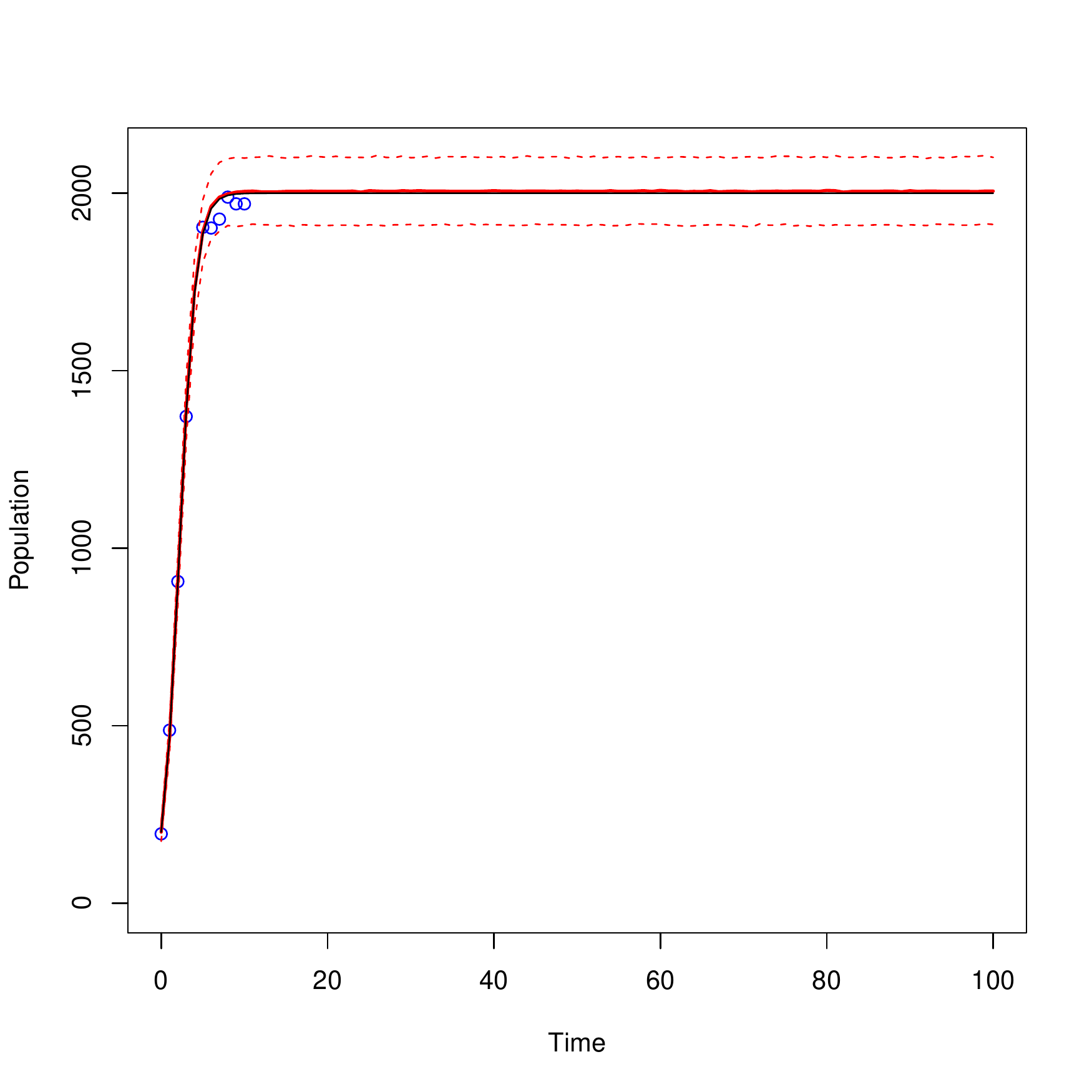}
}
\subfigure[]{
\includegraphics[width=1.96in]{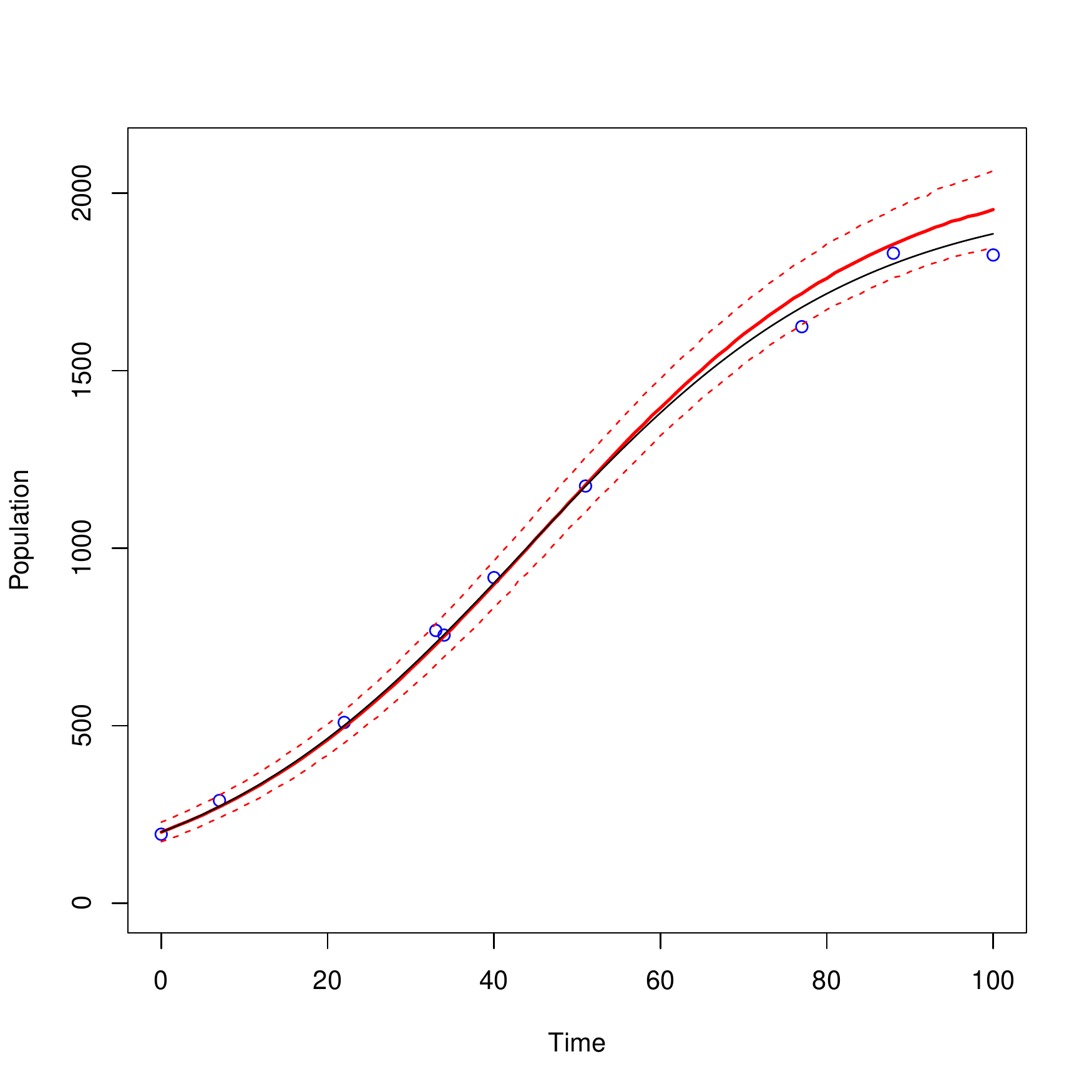}
}
\caption{10,000 MCMC samples are used to simulate $A_{\Phi}$ - optimal designs for (a) normal, (b) fast and (c) slow growth models. All models are simulated across 100 time points with a carrying capacity of 2000. The ground truth models are plotted in black. The optimal design points are plotted in blue and fit by a solid red curve, where the 2.5\% and 97.5\% quantiles of the predicted values are plotted as red dotted lines.}
\label{fig:SA1}
\end{figure}

\begin{figure}[H]
\centering
\subfigure[]{
\includegraphics[width=1.96in]{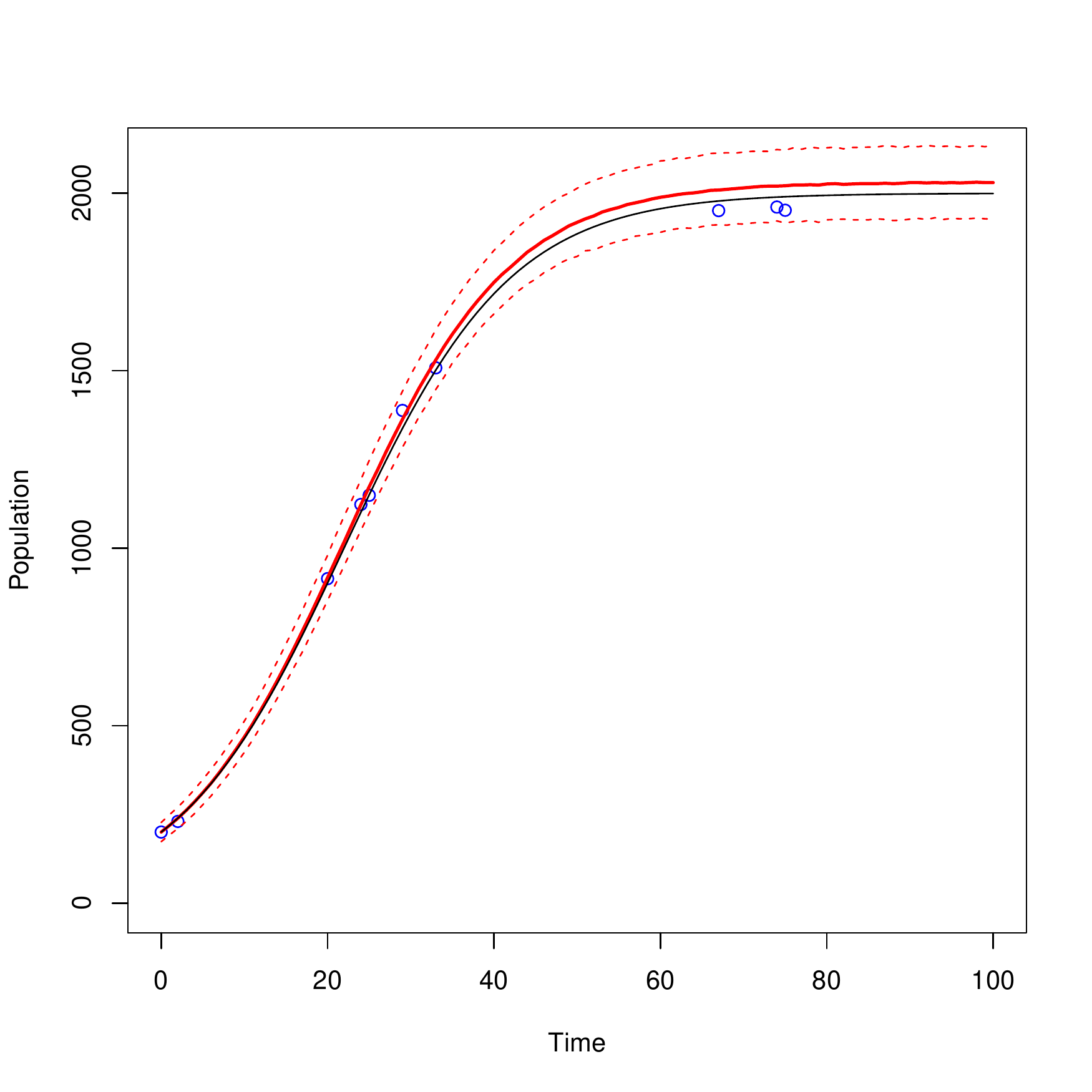}
}
\subfigure[]{
\includegraphics[width=1.96in]{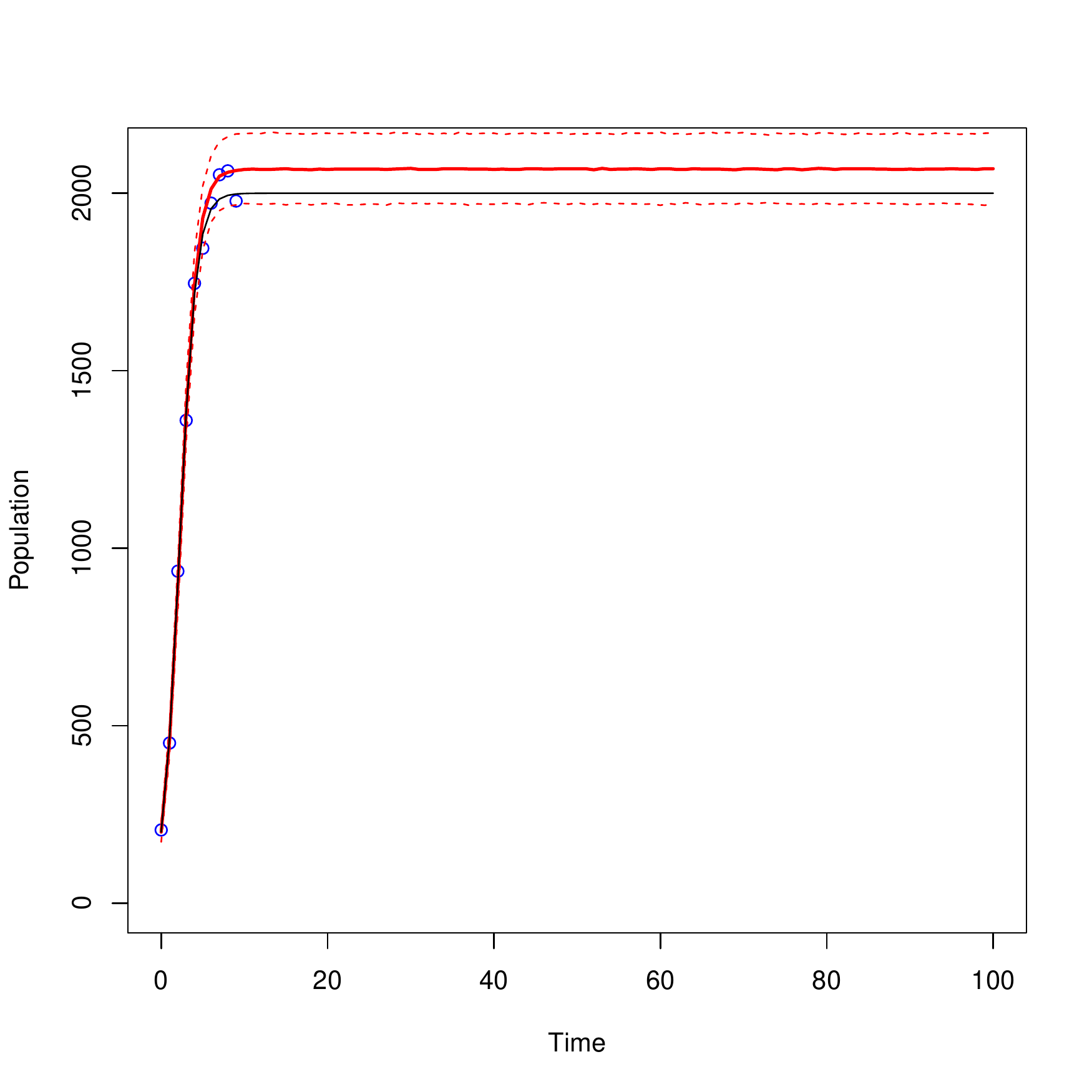}
}
\subfigure[]{
\includegraphics[width=1.96in]{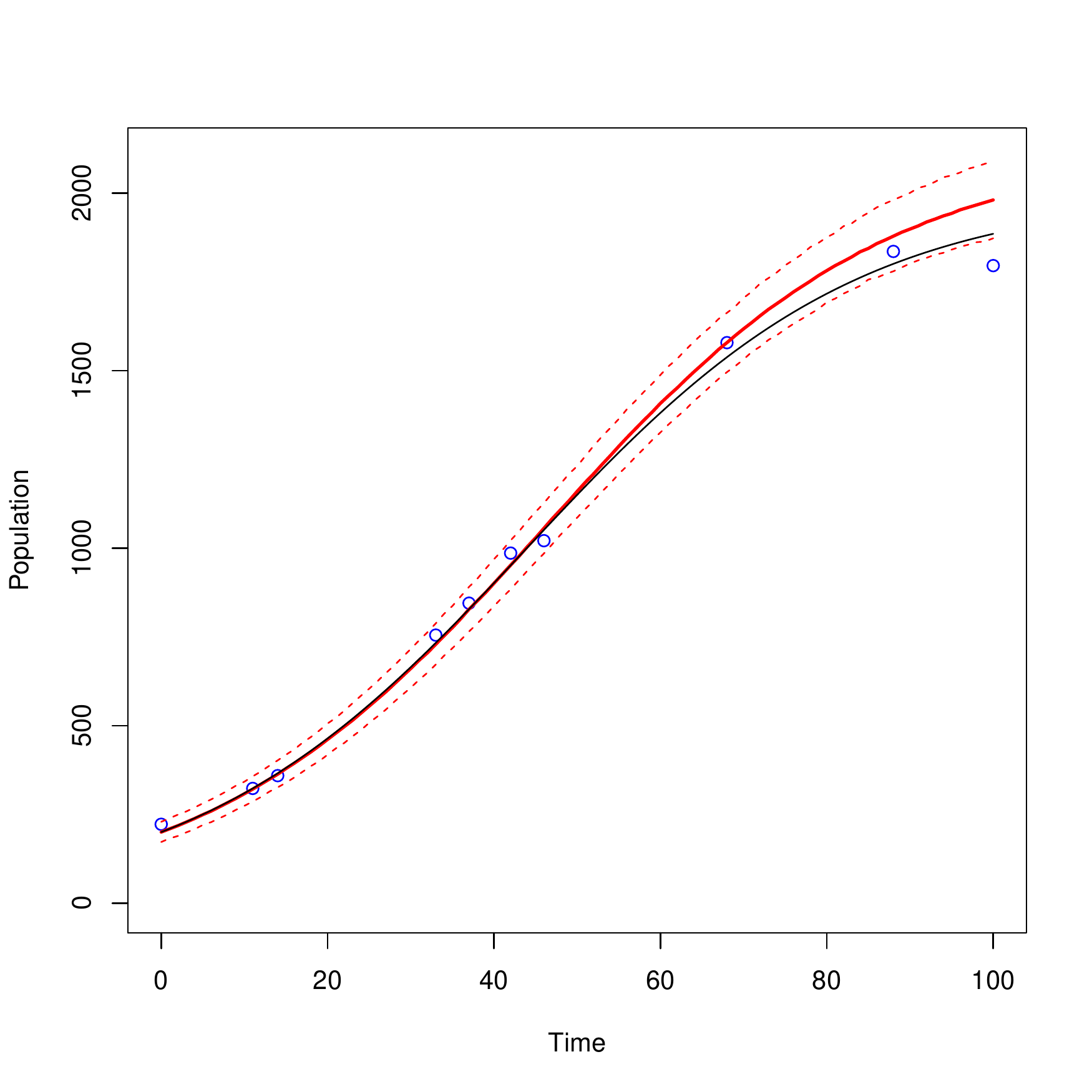}
}
\caption{10,000 MCMC samples are used to simulate $D_{\Phi}$ - optimal designs for (a) normal, (b) fast and (c) slow growth models. All models are simulated across 100 time points with a carrying capacity of 2000. The ground truth models are plotted in black. The optimal design points are plotted in blue and fit by a solid red curve, where the 2.5\% and 97.5\% quantiles of the predicted values are plotted as red dotted lines.}
\label{fig:SA2}
\end{figure}

\begin{figure}[H]
\centering
\subfigure[]{
\includegraphics[width=1.96in]{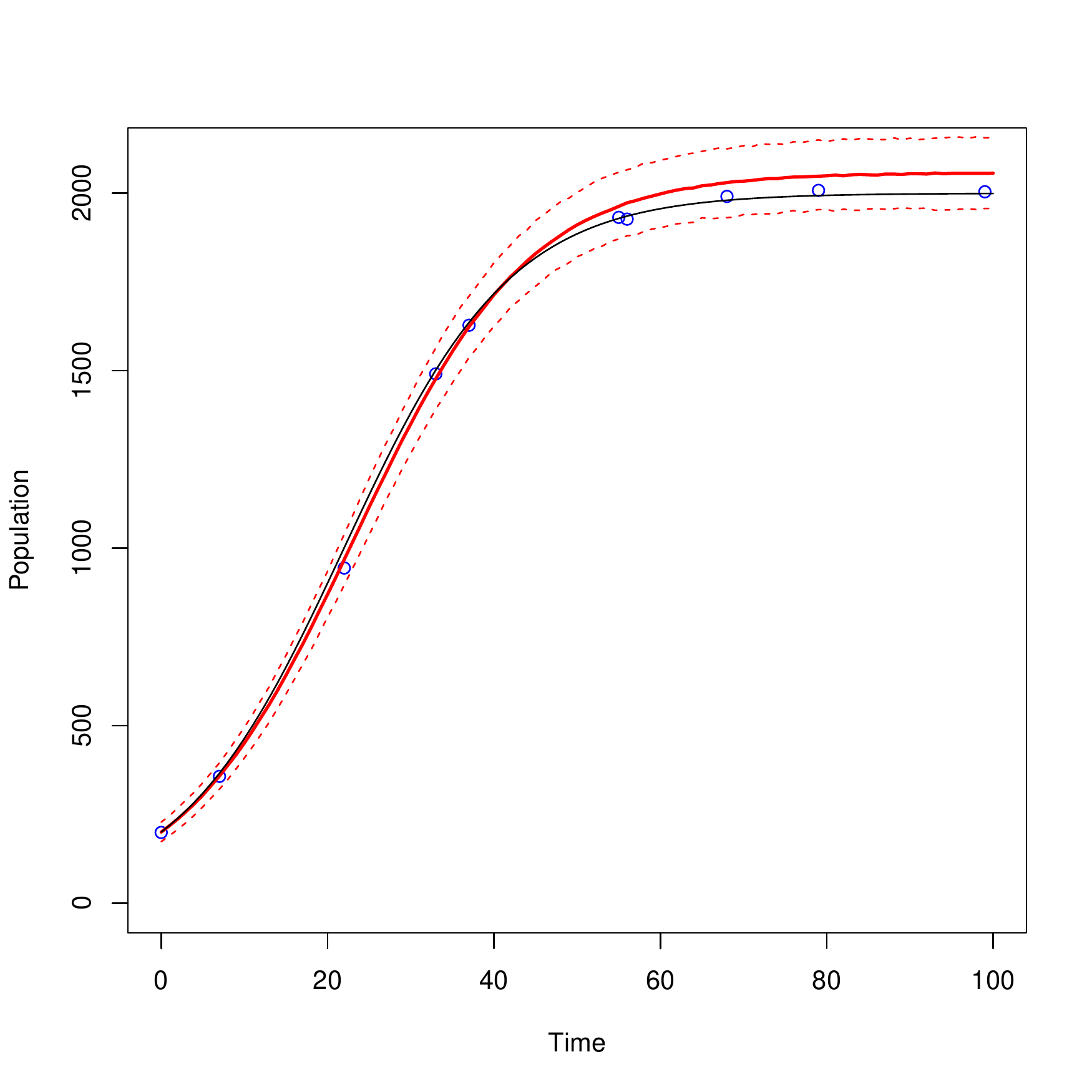}
}
\subfigure[]{
\includegraphics[width=1.96in]{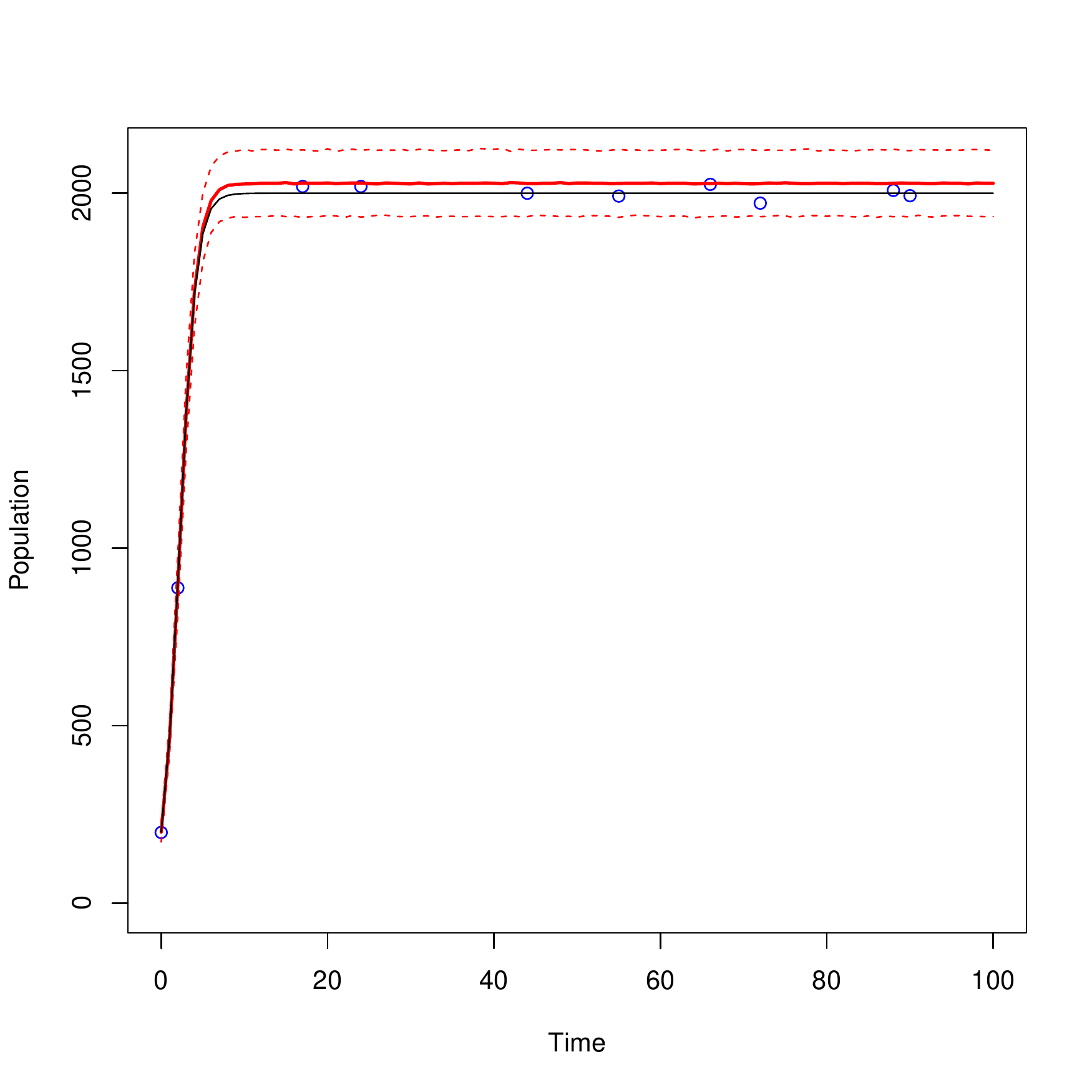}
}
\subfigure[]{
\includegraphics[width=1.96in]{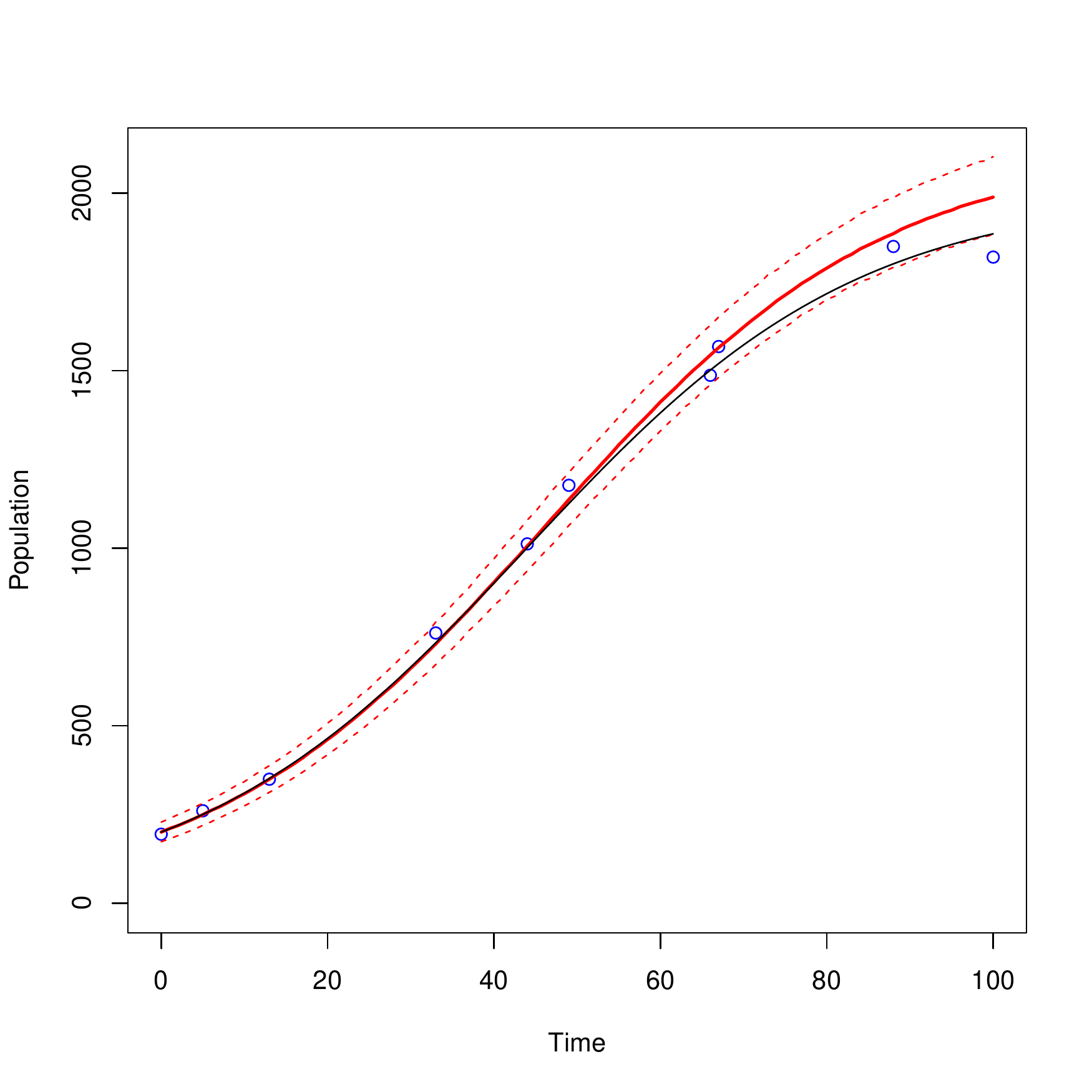}
}
\caption{10,000 MCMC samples are used to simulate I-optimal designs for (a) normal, (b) fast and (c) slow growth models. All models are simulated across 100 time points with a carrying capacity of 2000. The ground truth models are plotted in black. The optimal design points are plotted in blue and fit by a solid red curve, where the 2.5\% and 97.5\% quantiles of the predicted values are plotted as red dotted lines.}
\label{fig:SA3}
\end{figure}
\vspace{-4mm}
\noindent Simulated annealing successfully captures the dynamics of all growth models using each criterion. Figure \ref{fig:SA1} demonstrates the designs generated using $A_{\Phi}$ optimality criterion. Figure \ref{fig:SA2} fits the $D_{\Phi}$-optimal designs. Figure \ref{fig:SA3} illustrates the fit of I - optimal designs. In the next section, simulation studies are conducted using sequential optimality to demonstrate a new method for prediction based optimization. Sequential optimality is an adaptation of simulated annealing that predicts optimal design points.

\subsection{Sequential Design}

Sequential optimality searches subsets of the design space to find the next best point. Normal, fast and slow growth at 10\%, 100\% and 5\% rates respectively are simulated. All simulations use 10,000 MCMC samples plotted across 100 time points. The carrying capacity is 2000 specified as prior knowledge in equation (\ref{eq:prior}). The design point budget consists of ten points, and the algorithm uses the optimality criteria from equation (\ref{eq:opt}). As each criterion guides the sequential optimality process, the final optimal designs are compared for each model. Again, the ground truth model is plotted as a black curve. The optimal design points are plotted as blue points fit by a solid red curve with the 2.5\% and 97.5\% quantiles of the predicted values plotted as red dotted lines. As the system updates, our goal is to fit the ground truth model as accurately as possible with the final optimal design. I, $A_{\Phi}$, and $D_{\Phi}$ optimal designs are compared across normal, fast, and slow growth rates.

\begin{figure}[H]
\centering
\subfigure[]{
\includegraphics[width=1.96in]{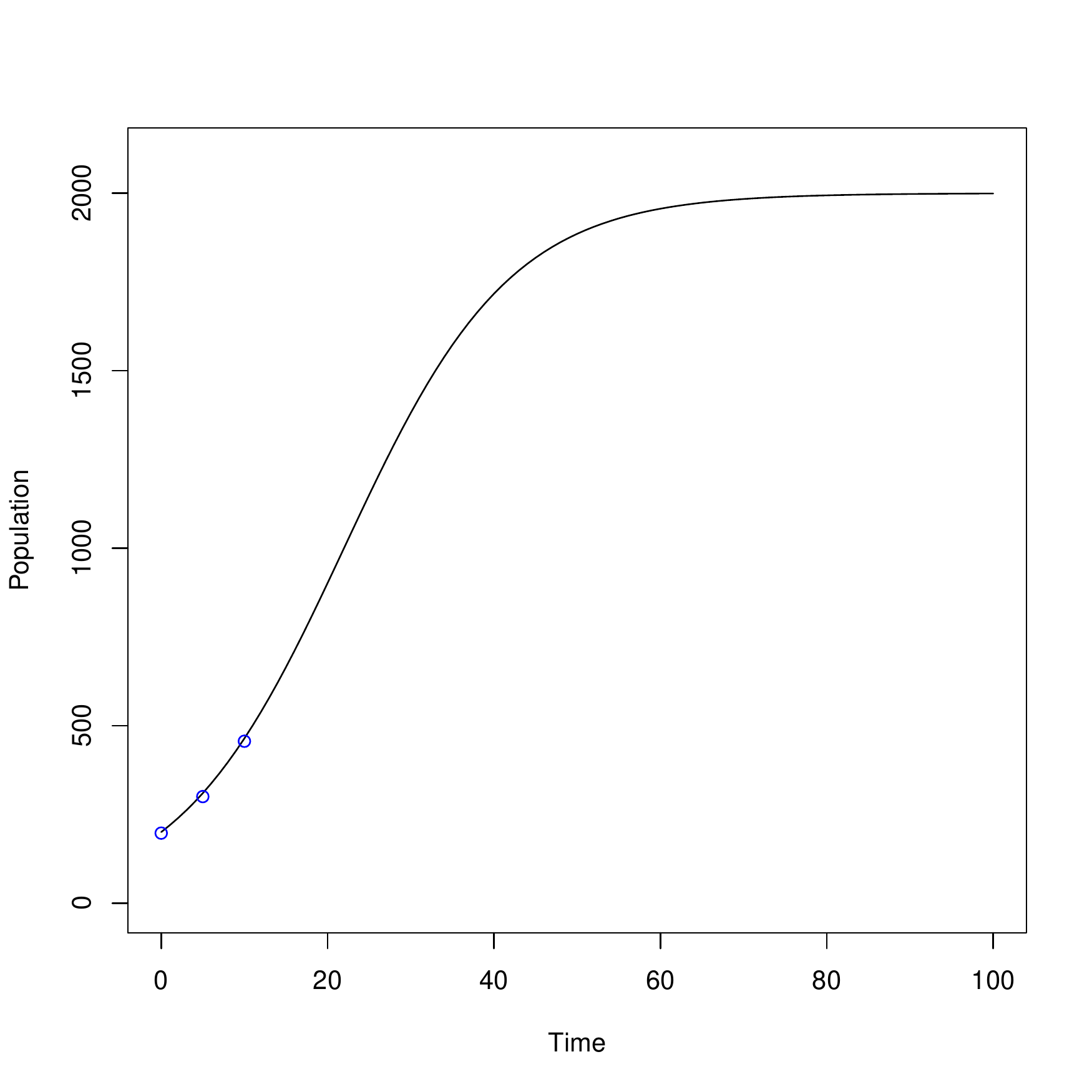}
}
\subfigure[]{
\includegraphics[width=1.96in]{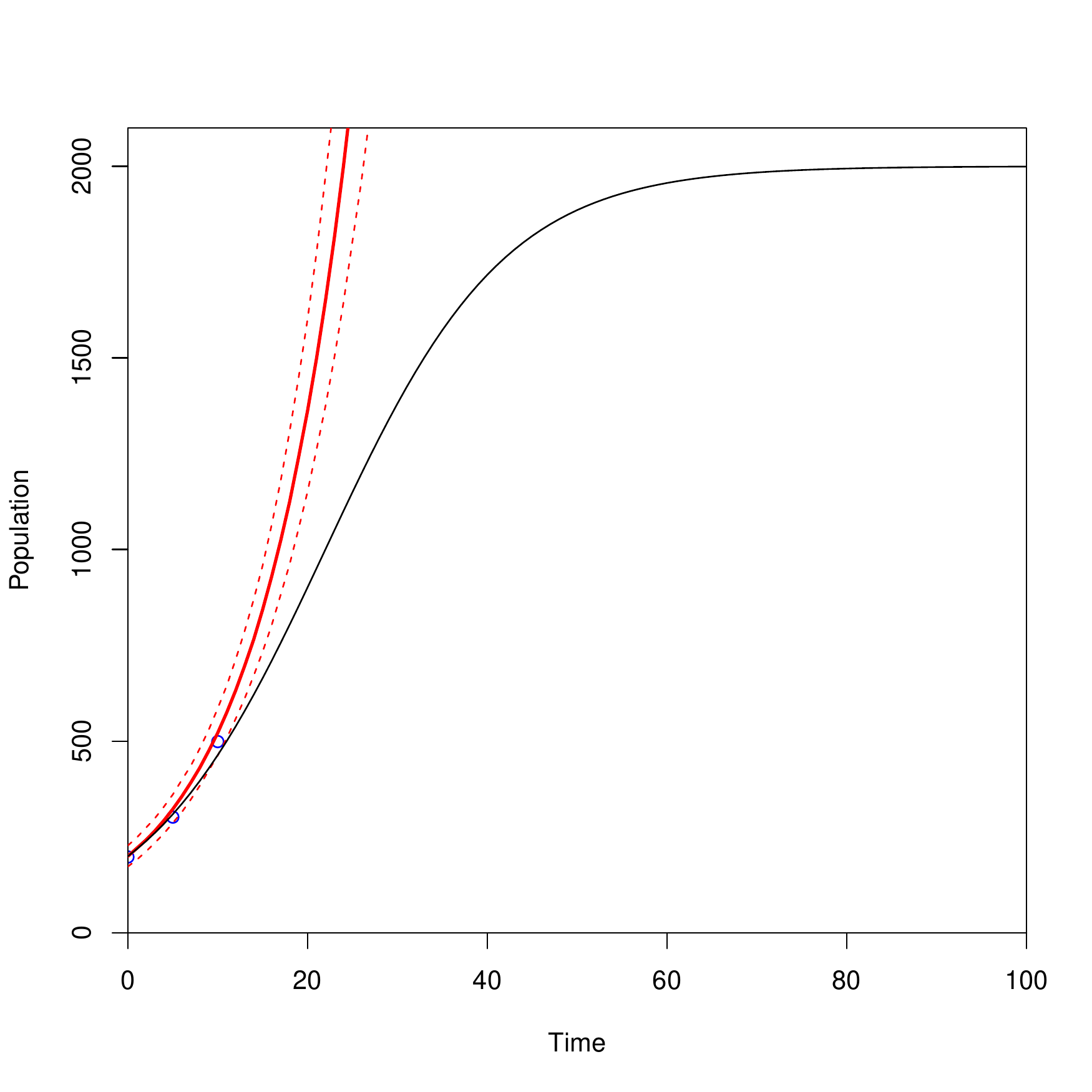}
}
\subfigure[]{
\includegraphics[width=1.96in]{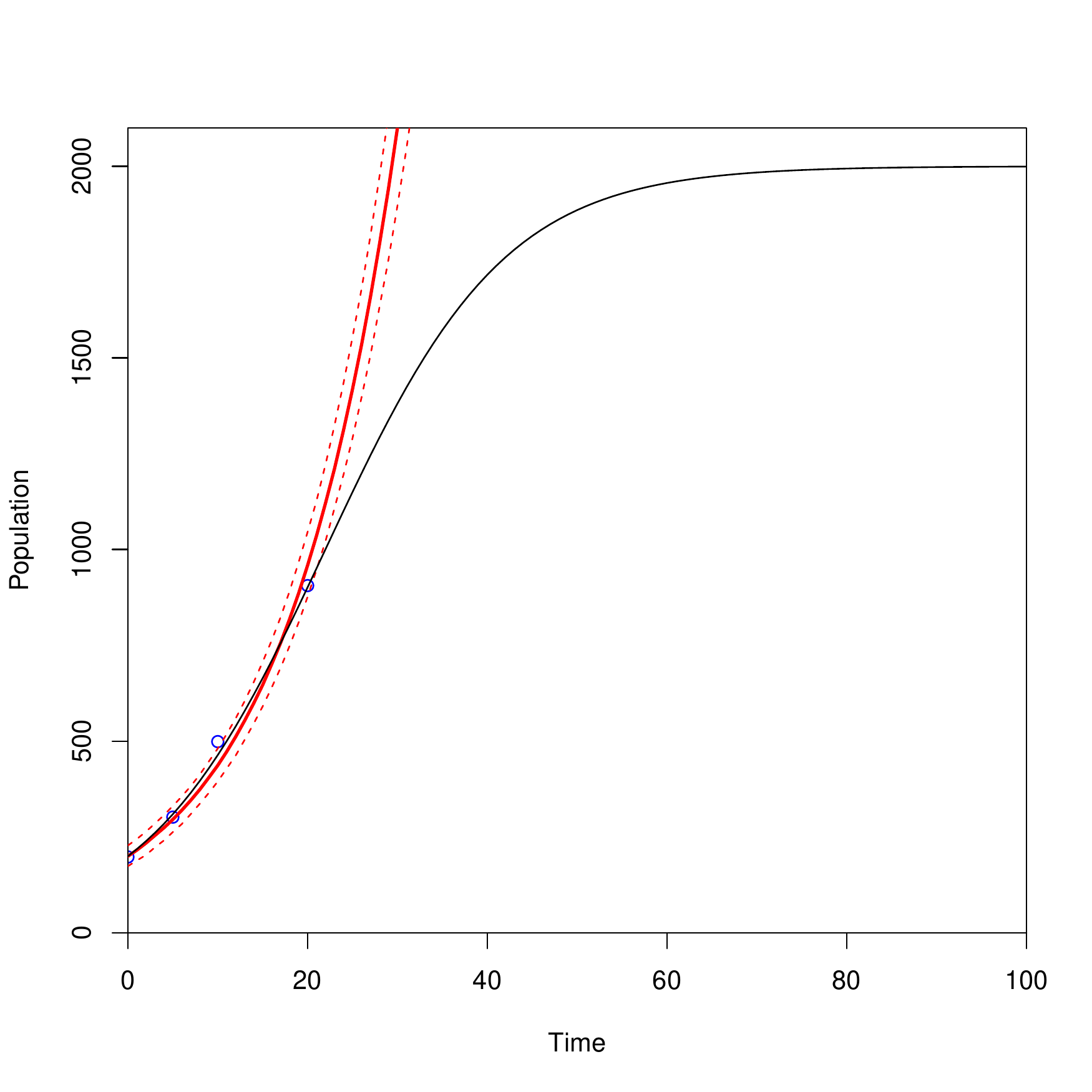}
}
\subfigure[]{
\includegraphics[width=1.96in]{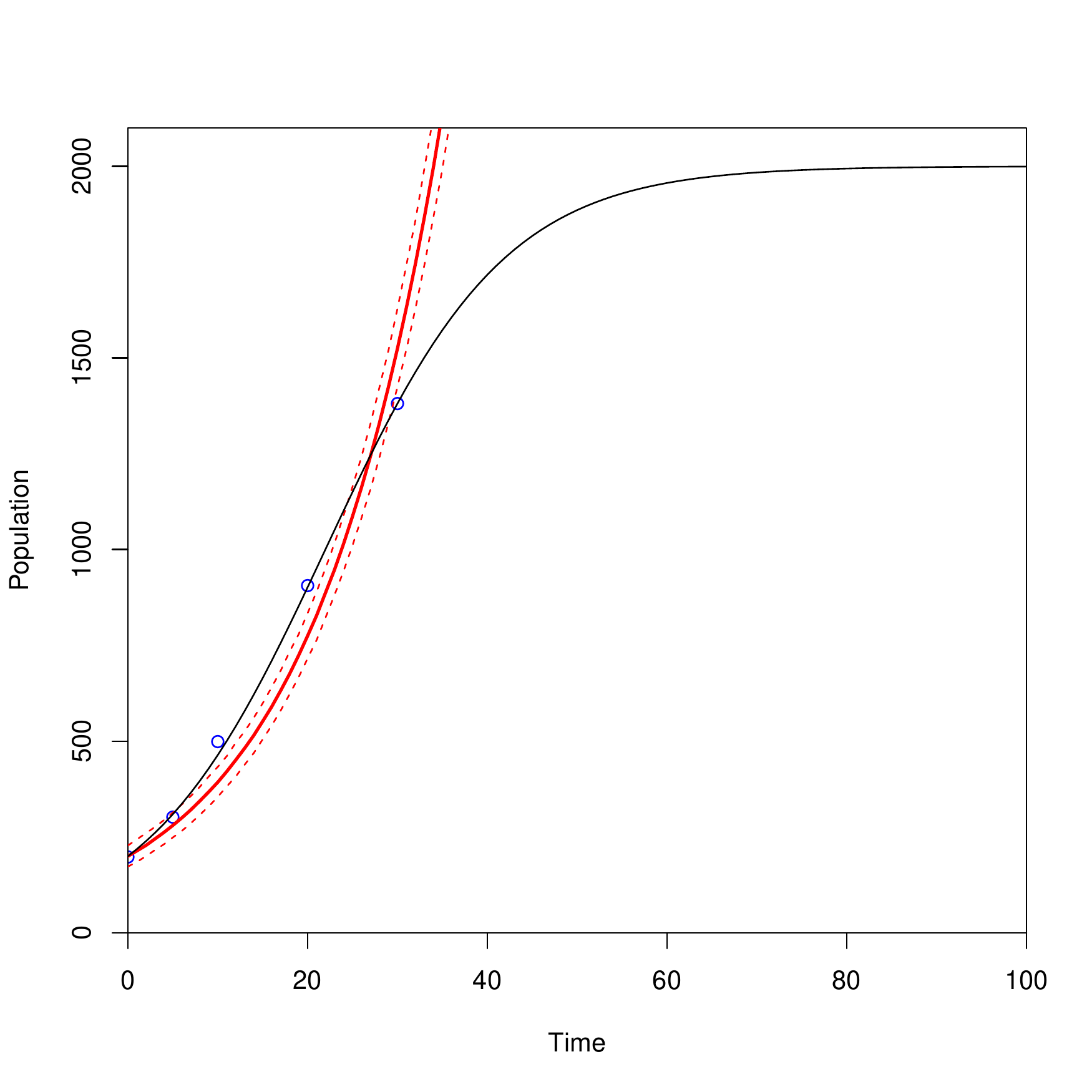}
}
\subfigure[]{
\includegraphics[width=1.96in]{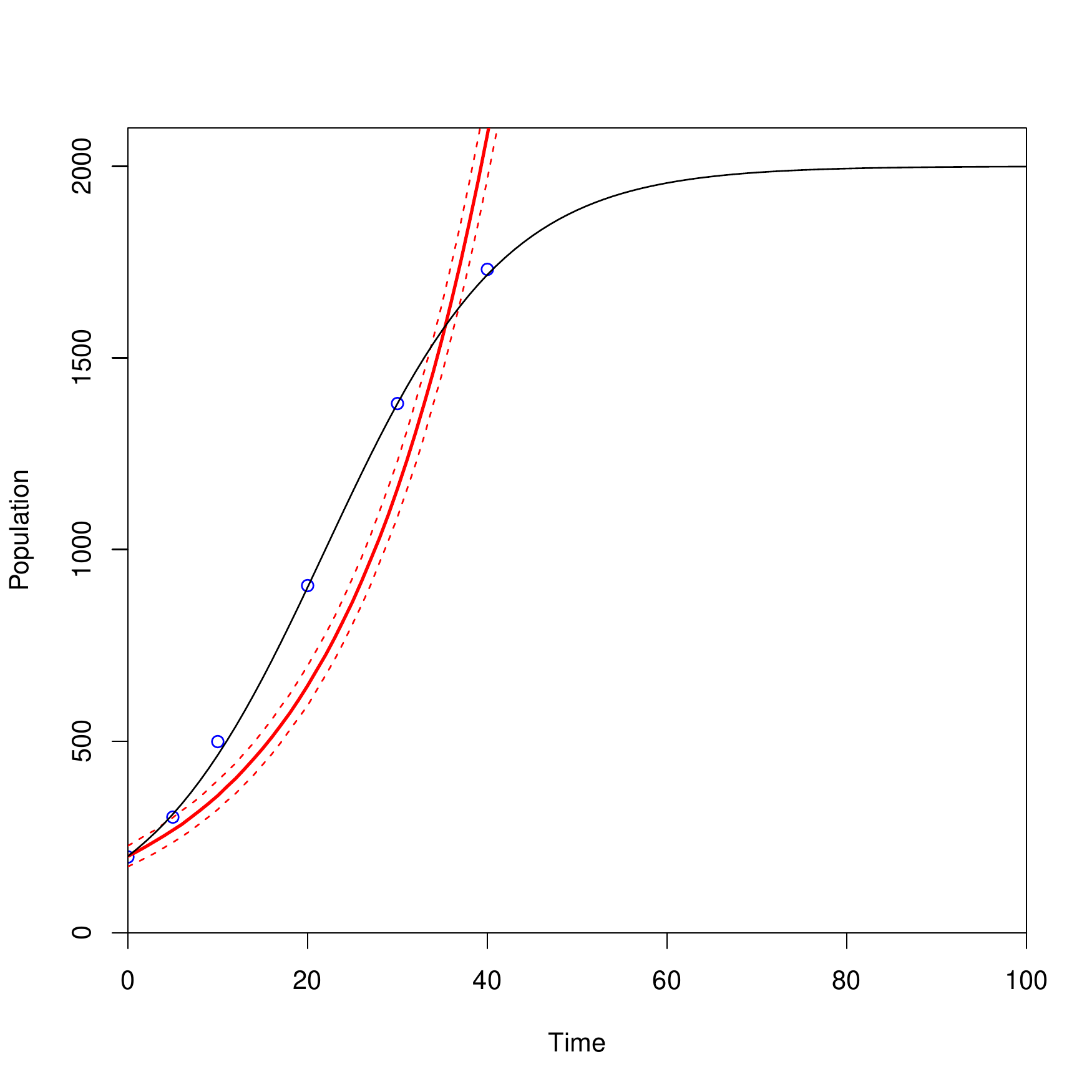}
}
\subfigure[]{
\includegraphics[width=1.96in]{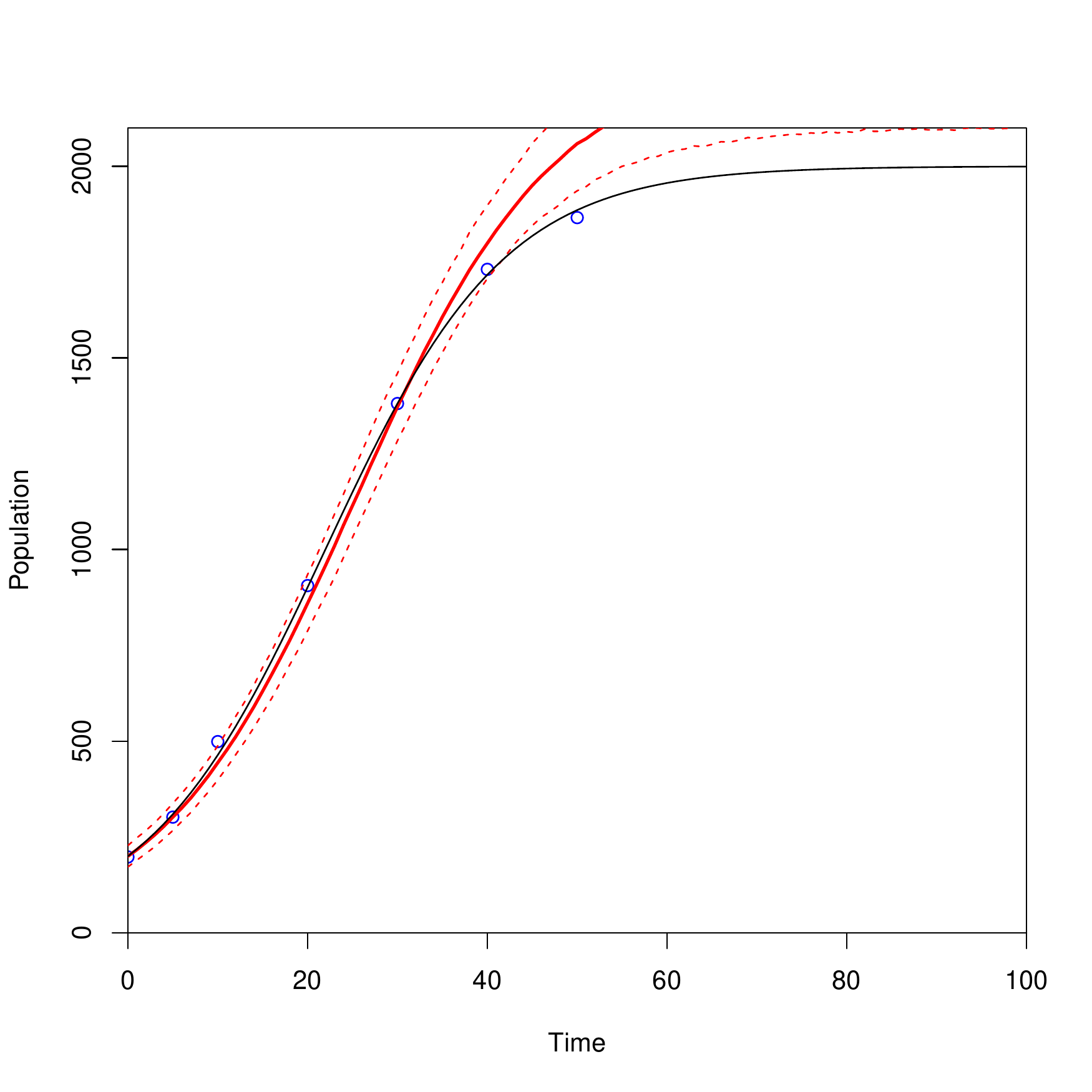}
}
\subfigure[]{
\includegraphics[width=1.96in]{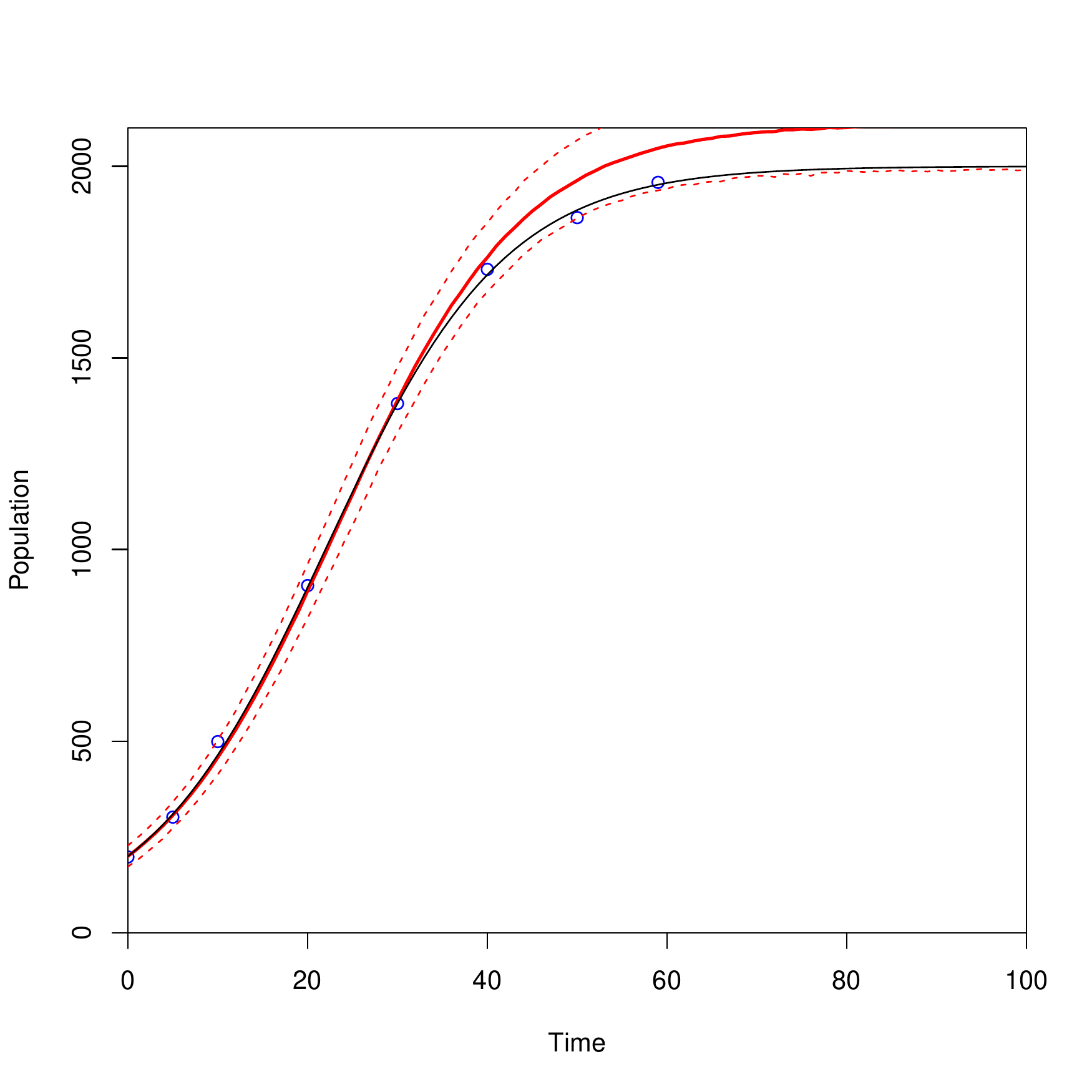}
}
\subfigure[]{
\includegraphics[width=1.96in]{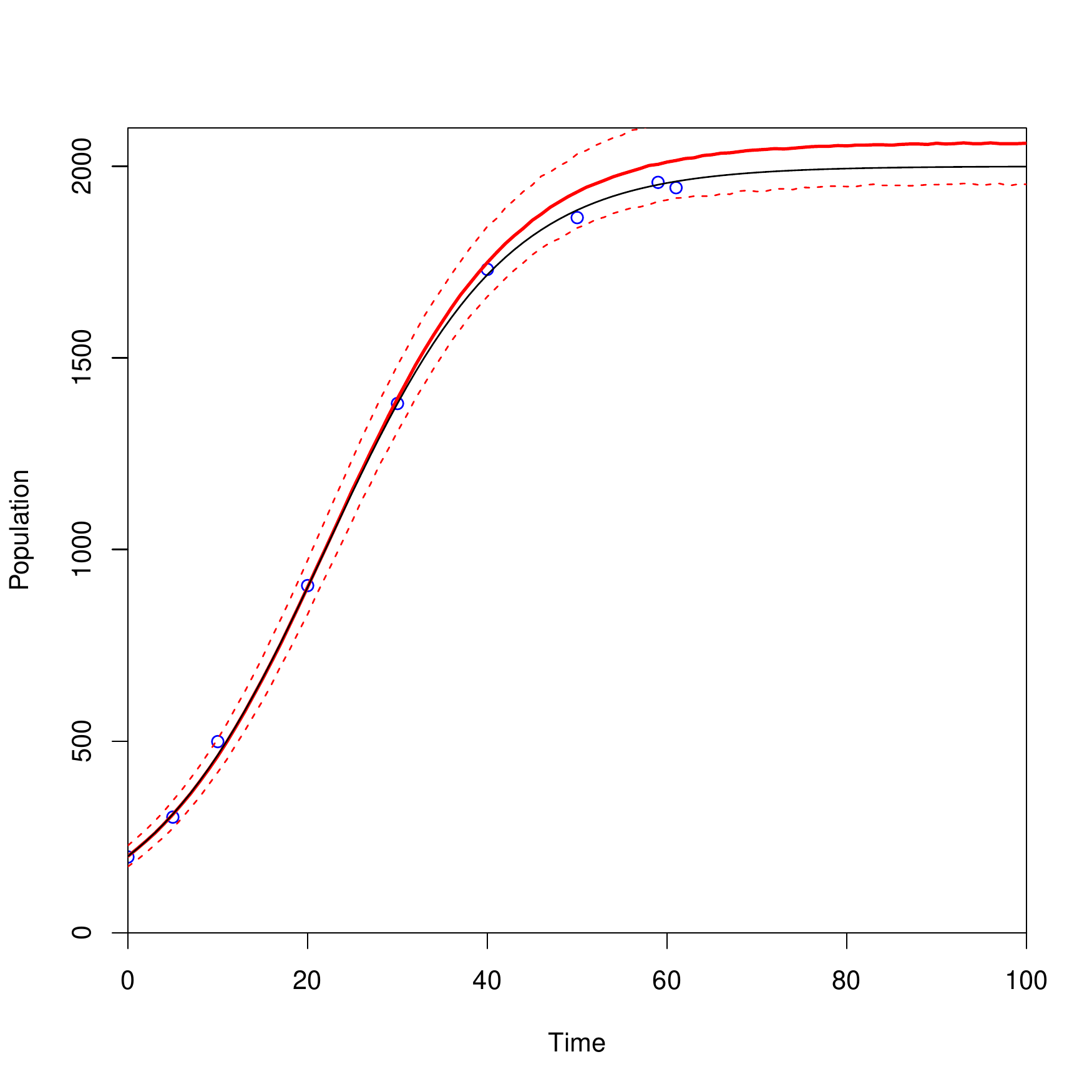}
}
\subfigure[]{
\includegraphics[width=1.96in]{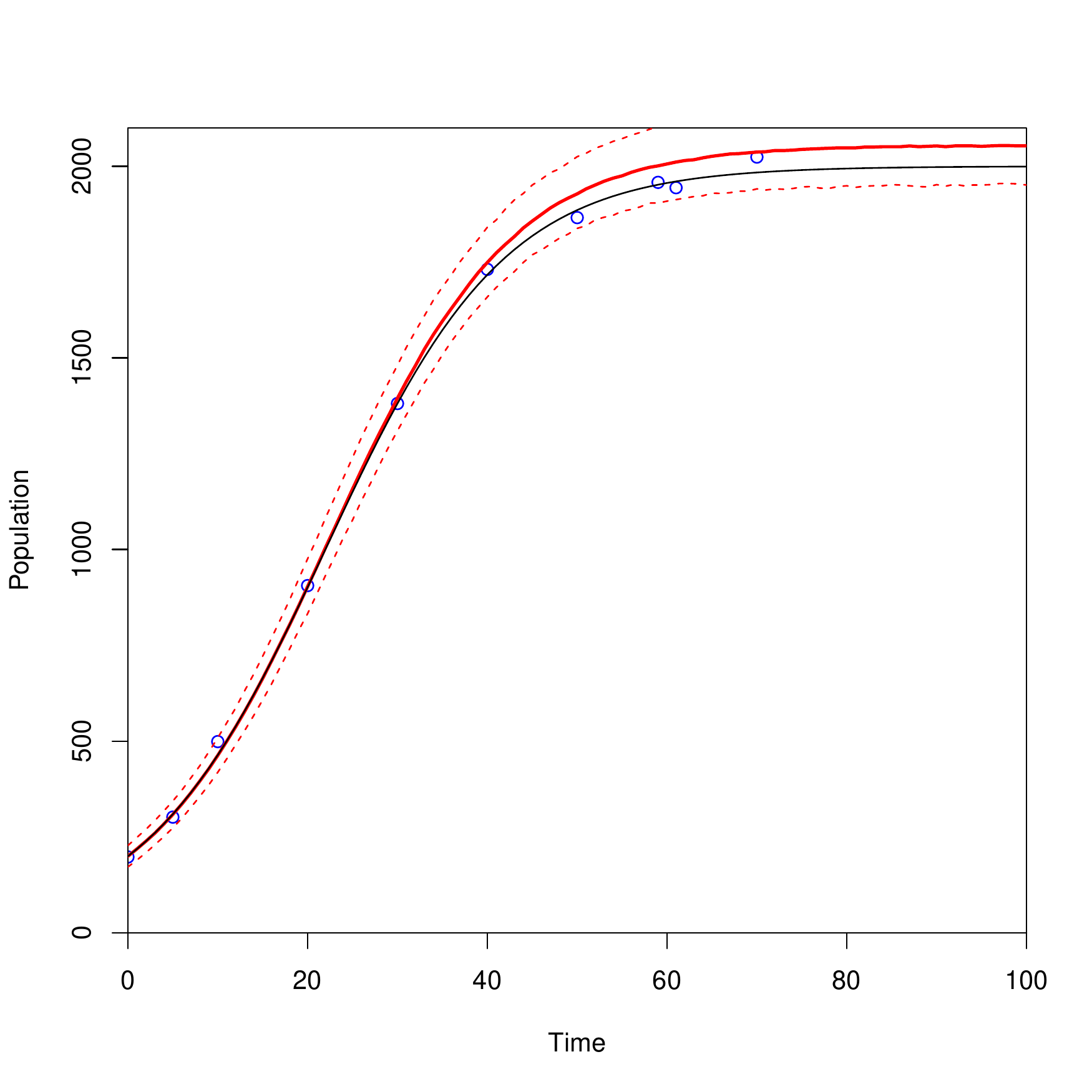}
}
\caption{10,000 MCMC samples are used to simulate the normal 10\% growth curve across a 100 day period with a carrying capacity of 2000. The ground truth model is plotted as a black curve. I-optimality criterion guides the sequential optimality process. The optimal design points are plotted in blue and are fit by a solid red curve. The red dotted lines represent the 2.5\% and 97.5\% quantiles of the predicted values. Panel (a) plots the base design of three points, panel (b) plots the fit of the base design, and panels (c)-(i) plot the fourth through tenth optimal points in the design as they are fit.}
\label{fig:SQ1}
\end{figure}

The simulations begin with the I-optimality criterion to demonstrate prediction based designs. Figure \ref{fig:SQ1} demonstrates sequential optimality using the prediction based I-optimality criterion. An initial design is set of three points given the logistic equation has two parameters, carrying capacity and growth rate. One hundred day seasons are modeled with a design point budget of ten. The ten candidate points following the base design are evaluated using the I-optimality criterion to select the point with the minimum prediction variance. Once the optimal point is added to the design, the process repeats until all ten design points are sampled. Based on the simulation, the prediction based I optimality criterion is successful in capturing the dynamics of a normal growth model. However, different combinations of criteria and growth models are simulated to test the robustness of the technique.

\begin{figure}[H]
\centering
\subfigure[]{
\includegraphics[width=1.96in]{Dyn3_EGCC1SimPred_10_2000_01_200_pic8.pdf}
}
\subfigure[]{
\includegraphics[width=1.96in]{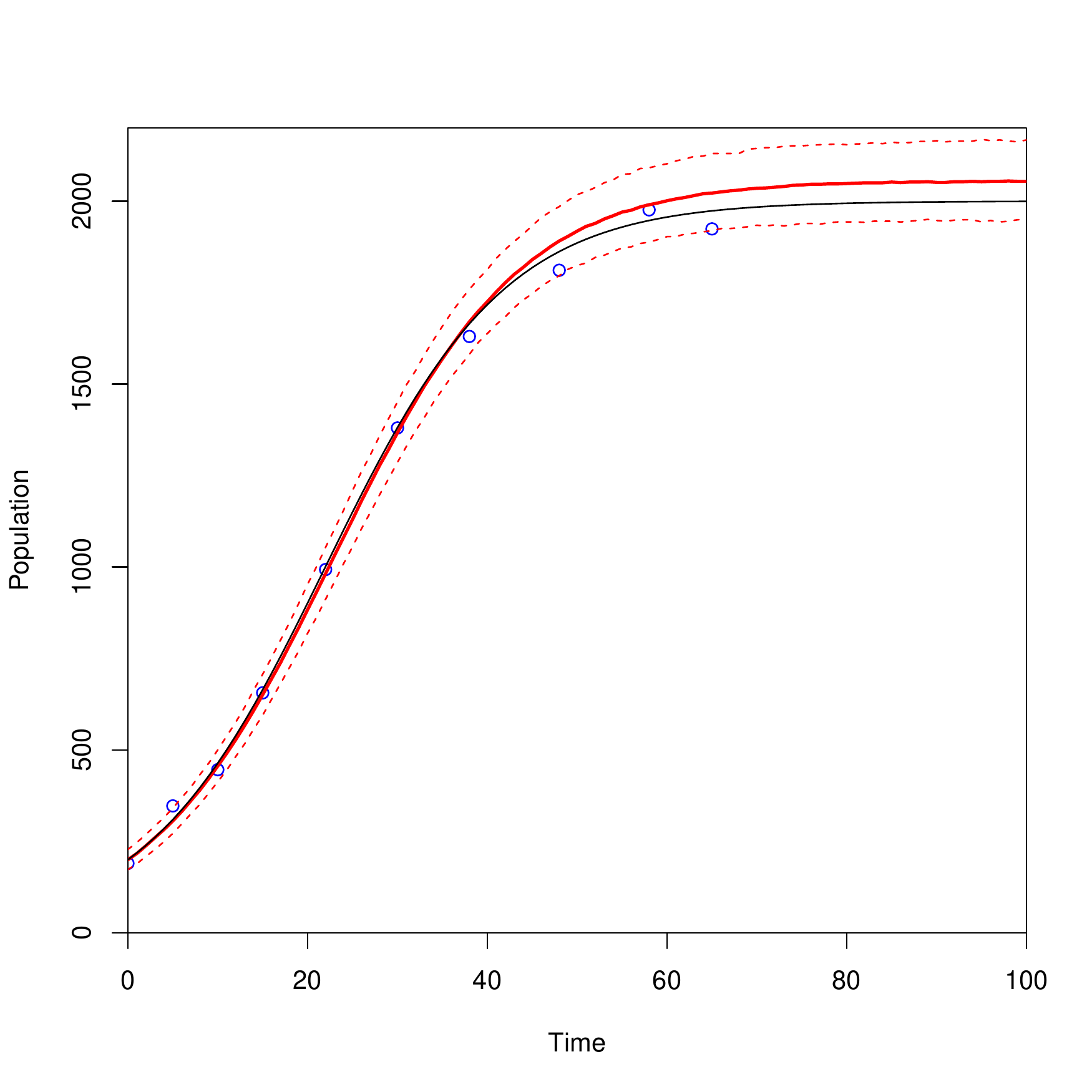}
}
\subfigure[]{
\includegraphics[width=1.96in]{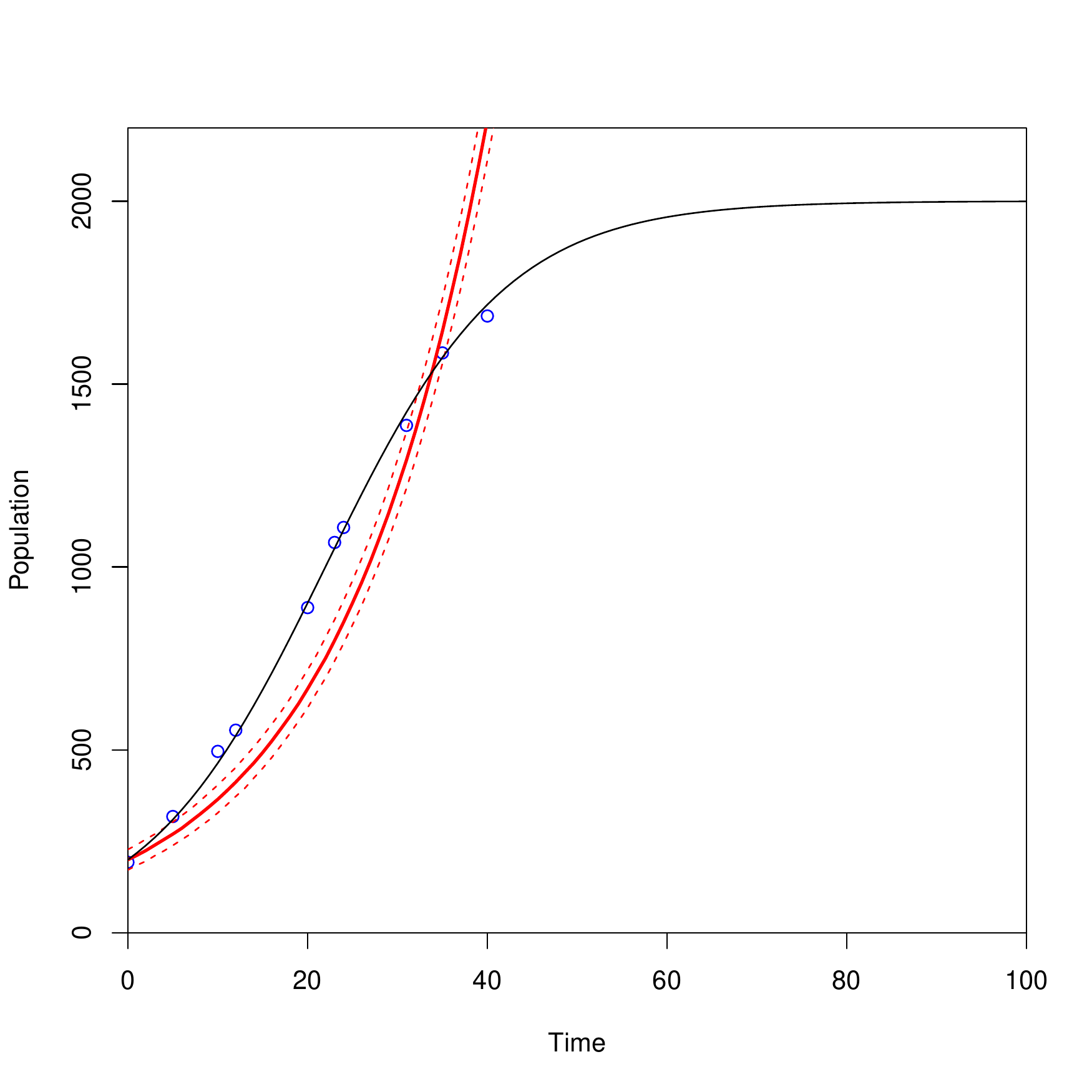}
}
\caption{10,000 MCMC samples are used in this simulation of a normal growth curve increasing at a 10\% rate across 100 time points with a carrying capacity of 2000. Sequential optimality is used to find the ten best points for (a) I, (b) $A_{\Phi}$ and (c) $D_{\Phi}$ optimal designs. The optimal points are plotted in blue and fit with a solid red curve. The red dotted lines represent the 2.5\% and 97.5\% quantiles of the predicted values of the model. The black curve represents the ground truth model.}
\label{fig:SQ2}
\end{figure}

\begin{figure}[h!]
\centering
\subfigure[]{
\includegraphics[width=1.96in]{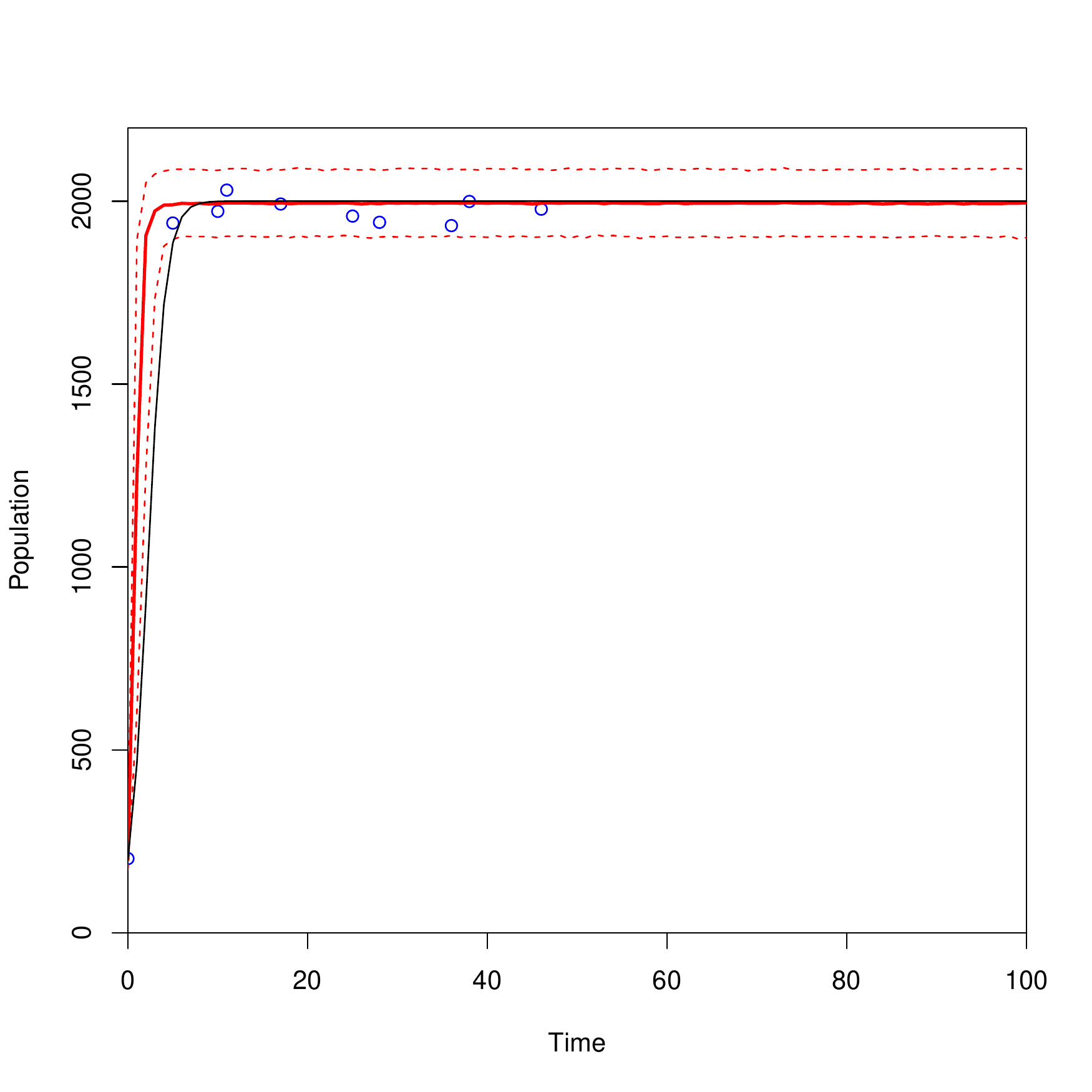}
}
\subfigure[]{
\includegraphics[width=1.96in]{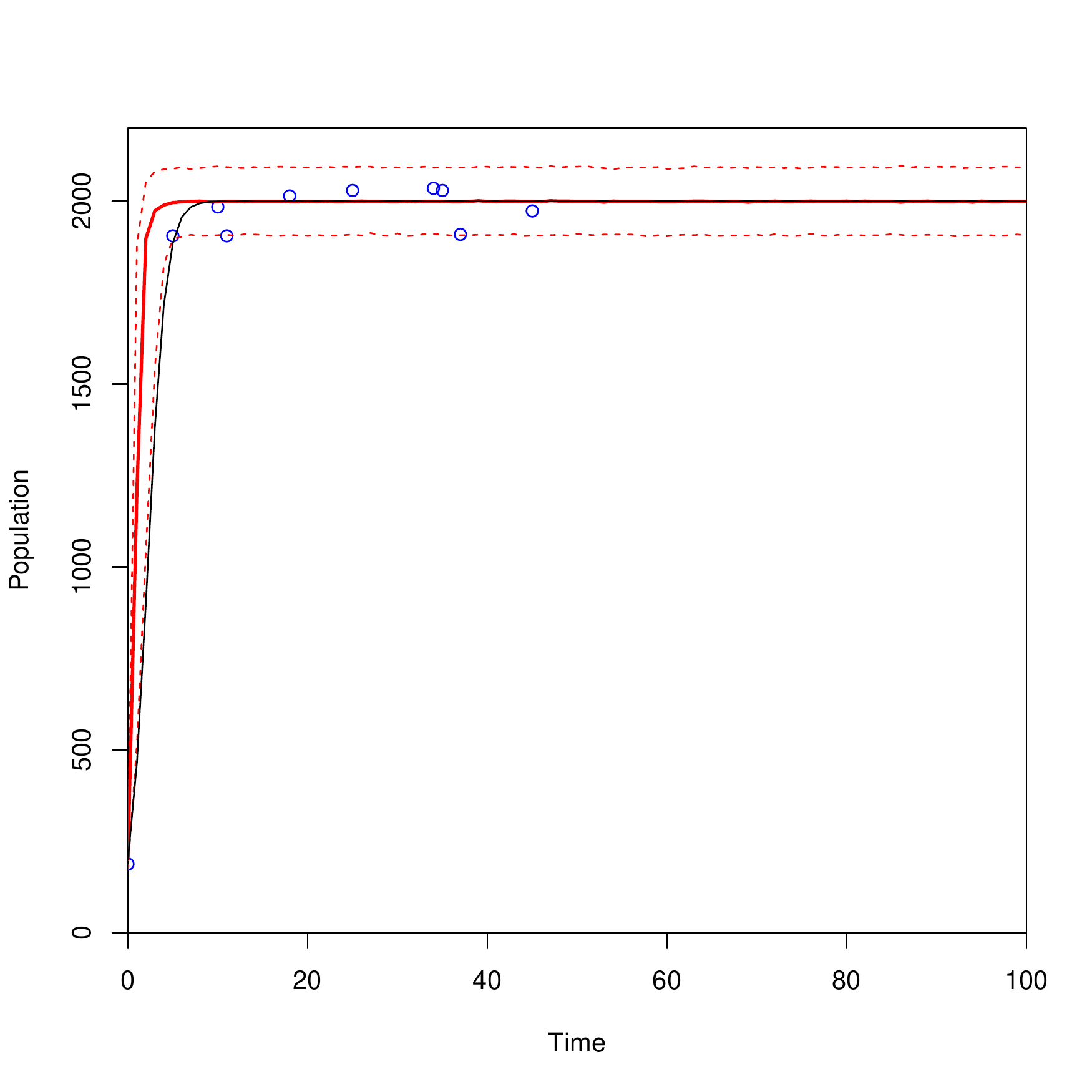}
}
\subfigure[]{
\includegraphics[width=1.96in]{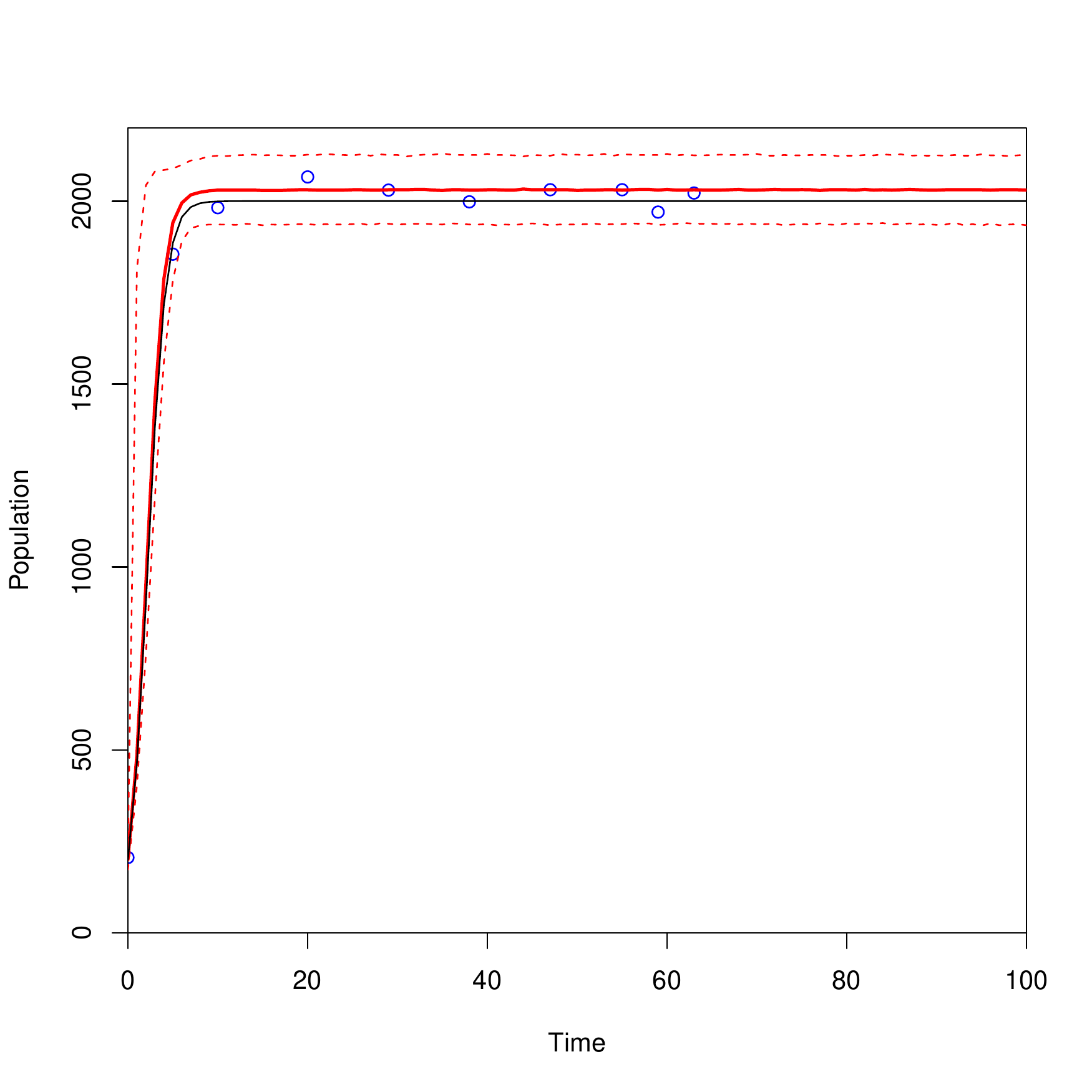}
}
\caption{10,000 MCMC samples are used in this simulation of a rapid growth curve increasing at a 100\% rate across 100 time points with a carrying capacity of 2000. Sequential optimality is used to find the ten best points for (a) I, (b) $A_{\Phi}$ and (c) $D_{\Phi}$ optimal designs. The optimal points are plotted in blue and fit with a solid red curve. The red dotted lines represent the 2.5\% and 97.5\% quantiles of the predicted values of the model. The black curve represents the ground truth model.}
\label{fig:SQ3}
\end{figure}

\begin{figure}[H]
\centering
\subfigure[]{
\includegraphics[width=1.96in]{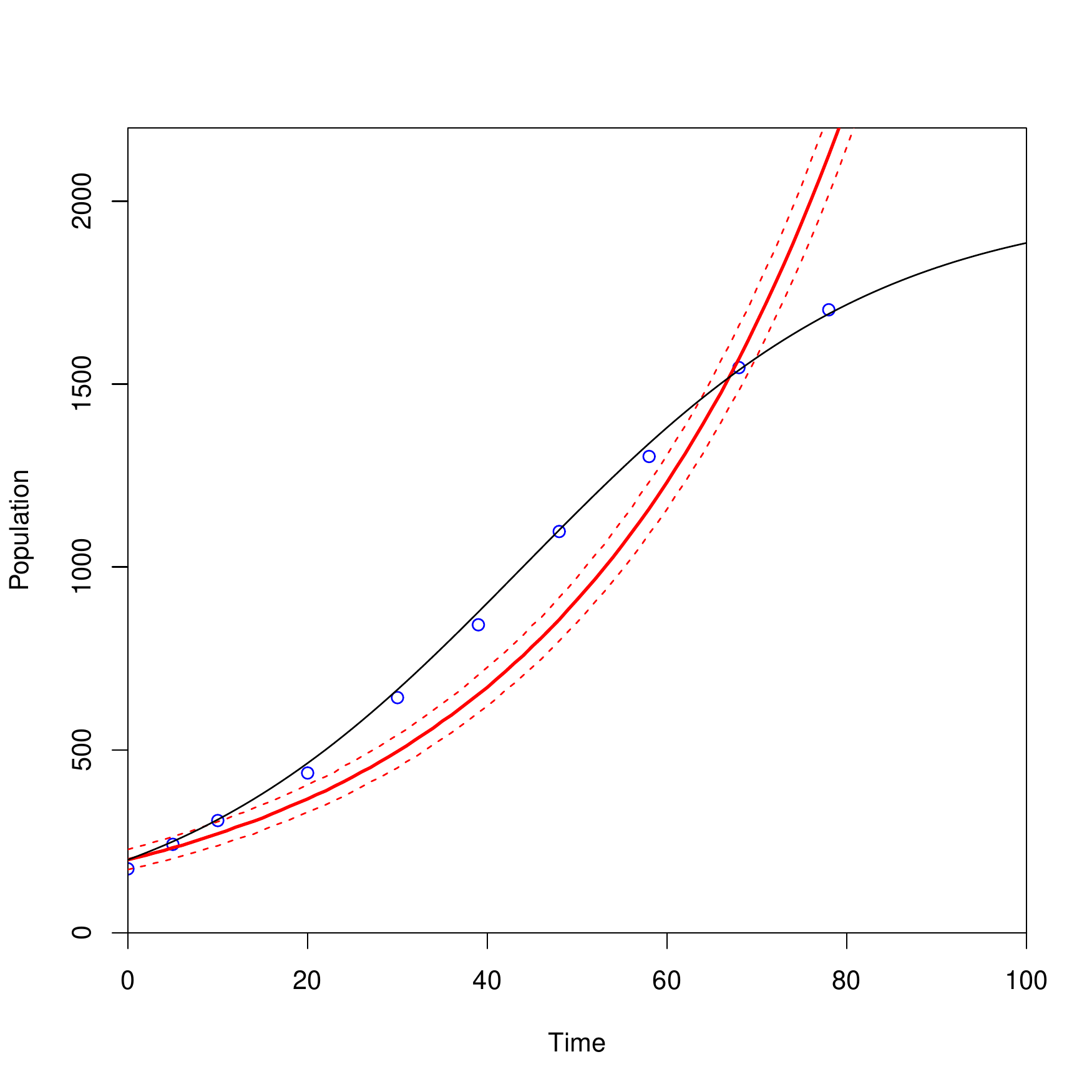}
}
\subfigure[]{
\includegraphics[width=1.96in]{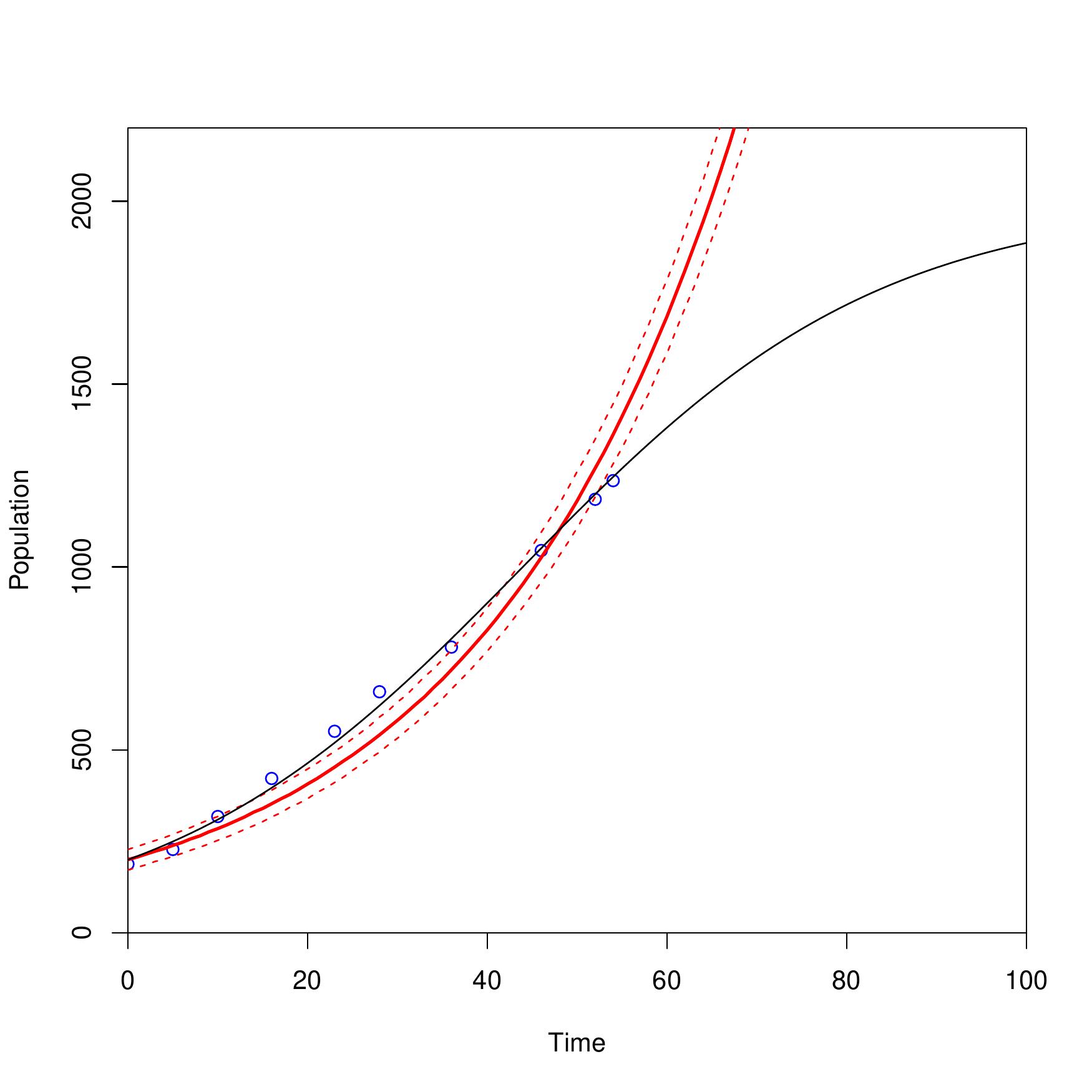}
}
\subfigure[]{
\includegraphics[width=1.96in]{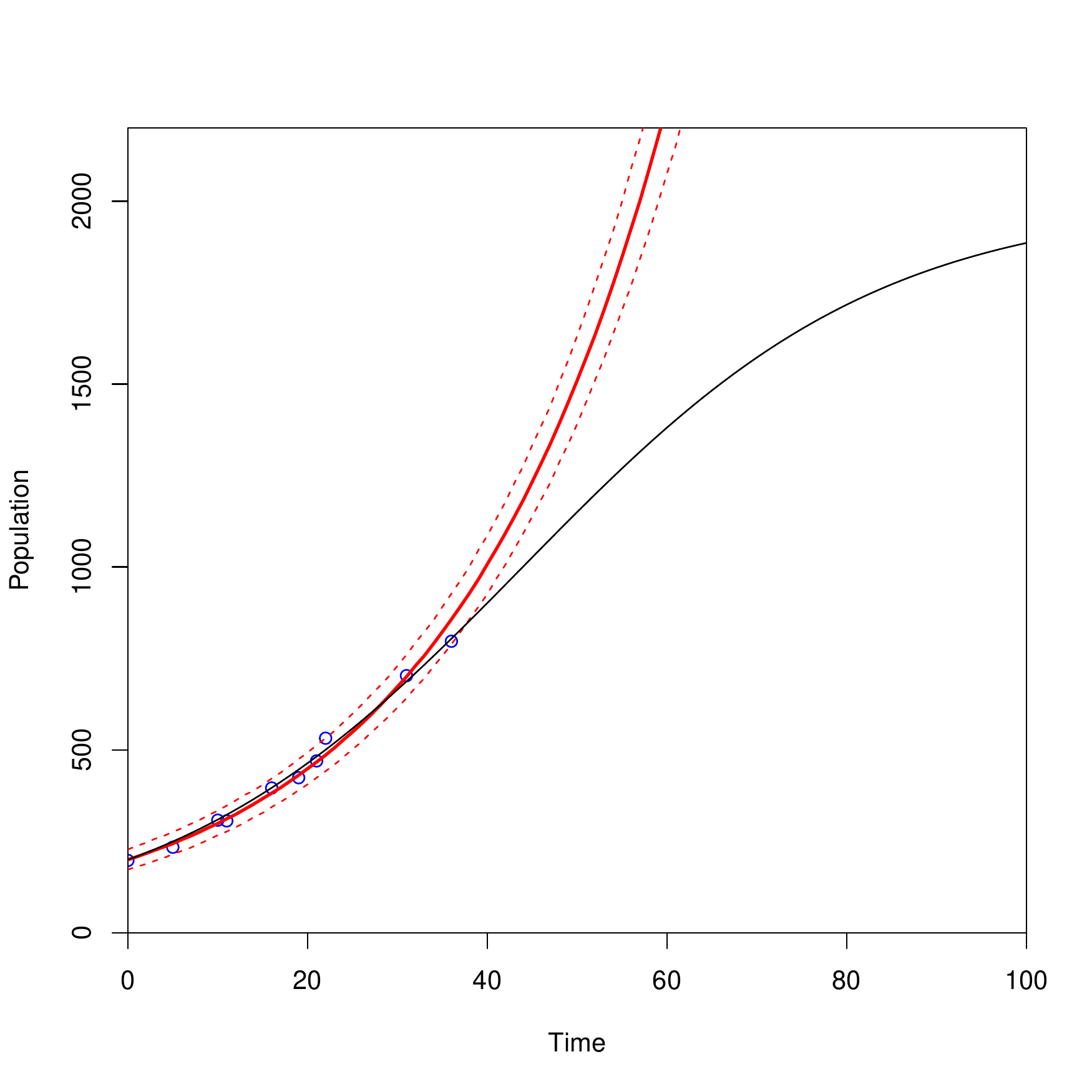}
}
\caption{10,000 MCMC samples are used in this simulation of a slow growth curve increasing at a 5\% rate across 100 time points with a carrying capacity of 2000. Sequential optimality is used to find the ten best points for (a) I, (b) $A_{\Phi}$ and (c) $D_{\Phi}$ optimal designs. The optimal points are plotted in blue and fit with a solid red curve. The red dotted lines represent the 2.5\% and 97.5\% quantiles of the predicted values of the model. The black curve represents the ground truth model.}
\label{fig:SQ4}
\end{figure}

The simulations of normal growth show varying in Figure \ref{fig:SQ2} results across the optimality criteria. The I and $A_{\Phi}$ optimal designs successfully capture the dynamics of the population with ten points. Whereas, the $D_{\Phi}$ optimal design did not capture the dynamic of the carrying capacity. This indicates the need for a larger design point budget or smaller design window when using $D_{\Phi}$ optimality criterion with a normal growth curve. On the other hand, all three optimality criteria capture the dynamics of the fast growth rate in Figure \ref{fig:SQ3}, which is expected given that the model plateaus rapidly. This implies that the design point budget could be decreased in this case. As for the slow growth rate model in Figure \ref{fig:SQ4}, no criteria could capture the population dynamics. This could indicate a need to increase the design point budget or decrease the design window for slow growth models. The size of the design point budget was set to ten points with a window of ten points to simply demonstrate this novel approach of sequential optimality.

Bayesian optimal designs \citep{Chaloner1995} are also examined for comparison purposes. Bayesian optimal designs can be written in the form of a utility function $U(d^{*}_{x_f})$, where $d^{*}_{x_f}$ represents a design chosen from the design region $d_{x_f}$. The design region is explored using Bayes I, D, and A optimal designs written mathematically as follows. 

Bayesian I-Optimal
\begin{equation}
\bar{U_I}(d^{*}_{x_f}) = min \int_{\Theta} min_{d_{x_f}} (Y_2 - E[Y_2|y_1,d_{x_f}])^2 p(Y_2 | y_1, d_{x_f}) d Y_2 \nonumber
\end{equation}

Bayesian D-Optimal
\begin{equation}
\bar{U_D}(d^{*}_{x_f}) = min \int_{\Theta} min_{d_{x_f}} |Cov(k,K | Y_2,y_1,d_{x_f})| p(Y_2 | y_1, d_{x_f}) d Y_2 \nonumber
\end{equation}

Bayesian A-Optimal
\begin{equation}
\bar{U_A}(d^{*}_{x_f}) = min \int_{\Theta} min_{d_{x_f}} tr(k,K | Y_2,y_1,d_{x_f}) p(Y_2 | y_1, d_{x_f}) d Y_2 \nonumber
\end{equation}

\noindent $Y_2$ represents the predicted design points, while $y_1$ represents the current design. $\Theta$ is the parameter space including parameters $k$ and $K$ from the model in equation (\ref{eq:1}). Similar to the criteria from equation (\ref{eq:opt}), the Bayesian optimality criteria are minimizing the posterior predictive distribution across the parameter space $\Theta$ and design space $d_{x_f}$. The Bayesian I-Optimal designs minimizes the distance between the squared predictions. Bayesian D-optimal designs minimize the determinant of the covariance of the posterior predictive distribution. The Bayesian A-optimal designs minimize the trace of the posterior predictive distribution. Implementing the Bayesian criteria into our process of sequential optimality gives various results. The frequency tables illustrate the precision of each criteria based on the probabilities assigned to each candidate point.

\begin{figure}[H]
\centering
\subfigure[]{
\includegraphics[width=1.96in]{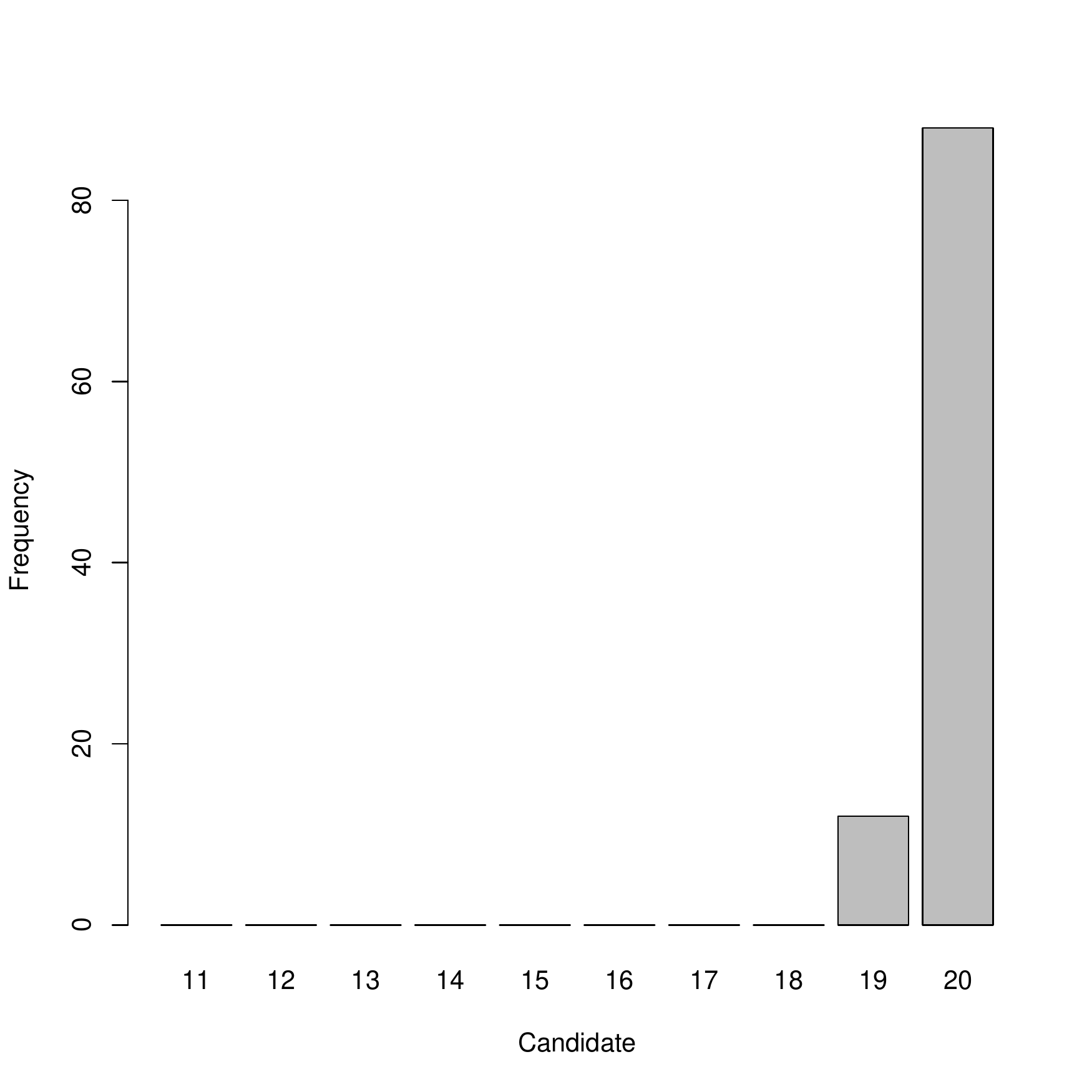}
}
\subfigure[]{
\includegraphics[width=1.96in]{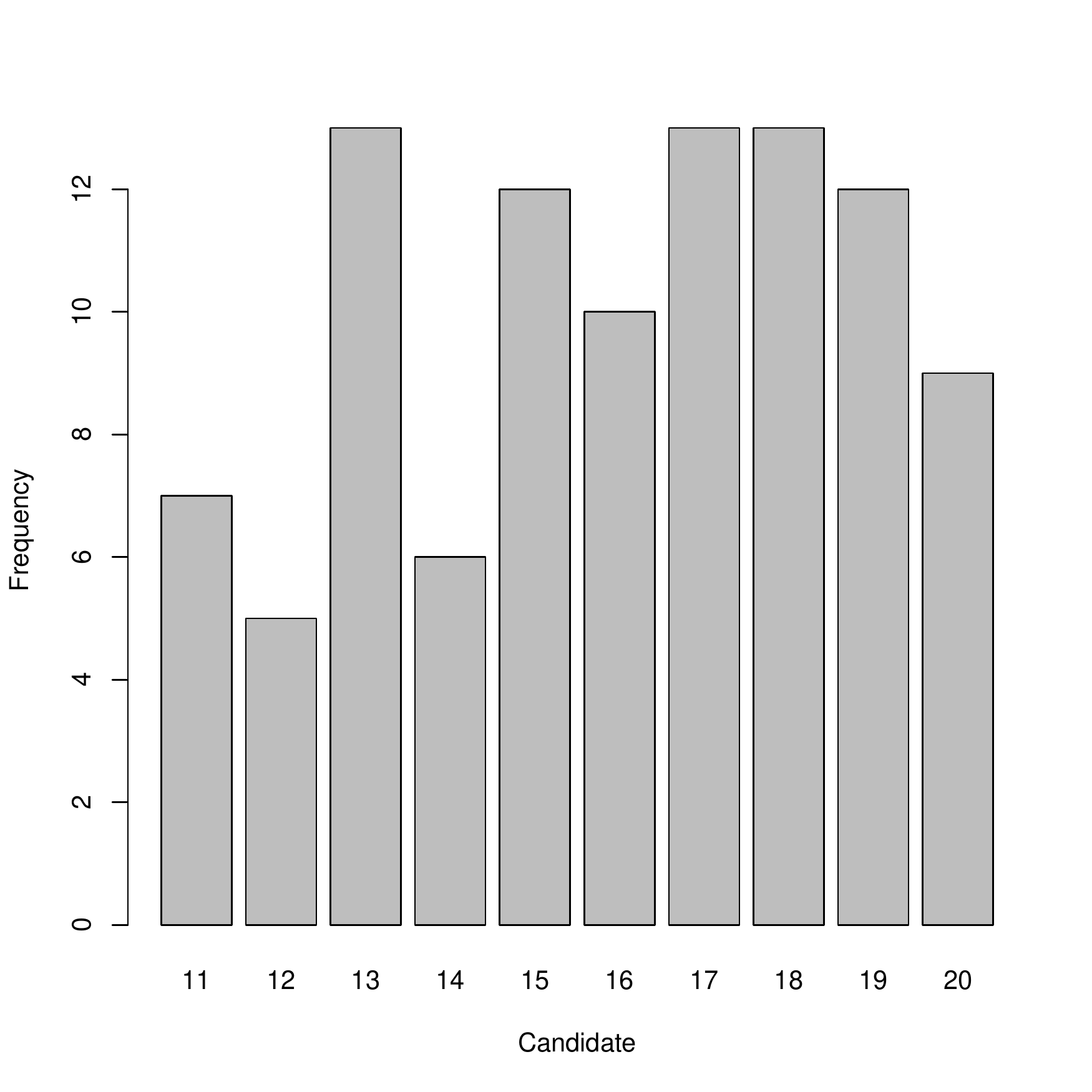}
}
\subfigure[]{
\includegraphics[width=1.96in]{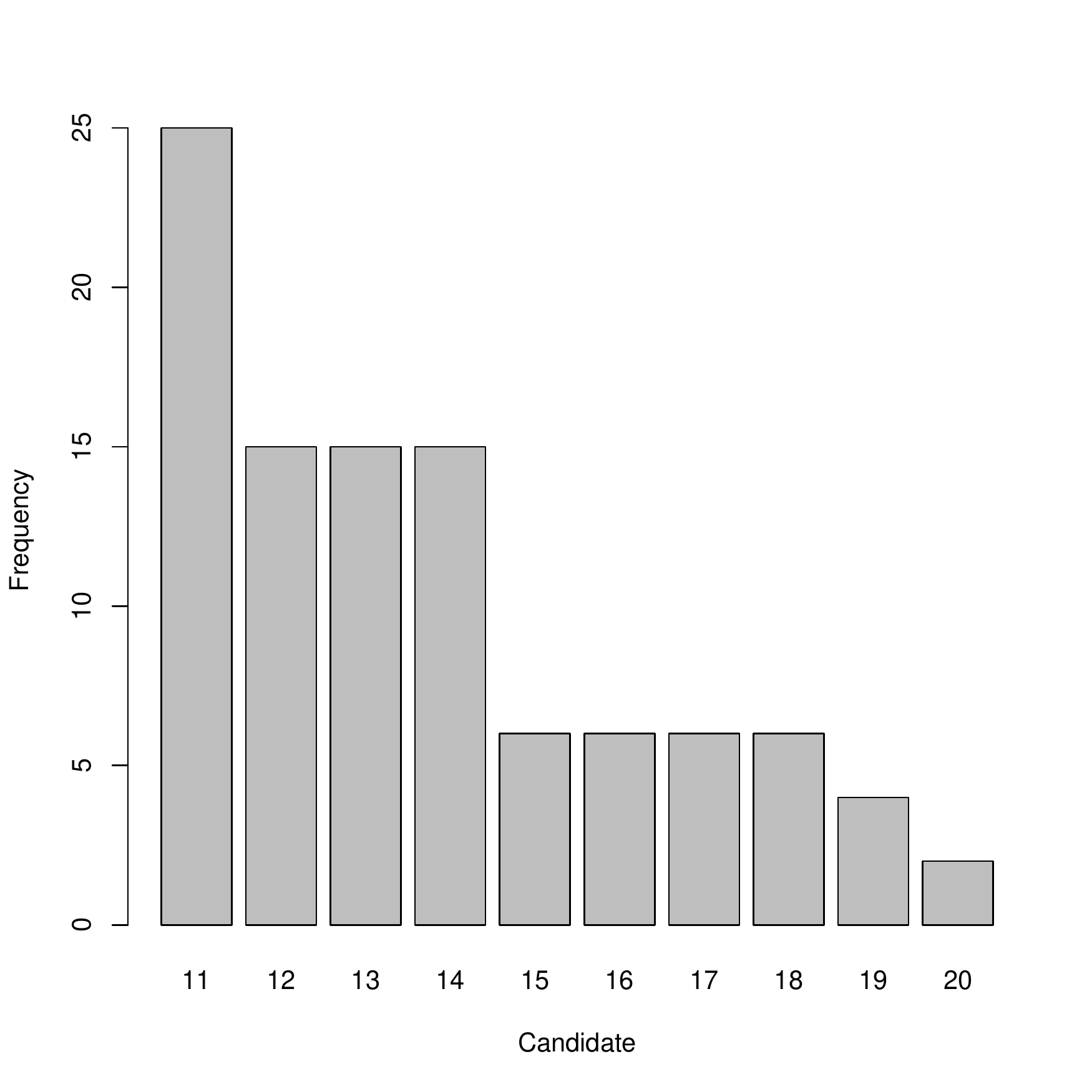}
}
\caption{Bayesian optimality criteria are used to guide the sequential optimality process. The frequency tables plot the probabilities assigned to the first ten candidate points following the initial base design for a normal growth model. Panel (a) represents candidate probabilities under the $\bar{U_I}$ criterion. Panel (b) plots probabilities of the candidates for $\bar{U_A}$ criterion. Panel (c) provides the frequencies associated with $\bar{U_D}$ criterion.}
\label{fig:SQ5}
\end{figure}

\begin{figure}[H]
\centering
\subfigure[]{
\includegraphics[width=1.96in]{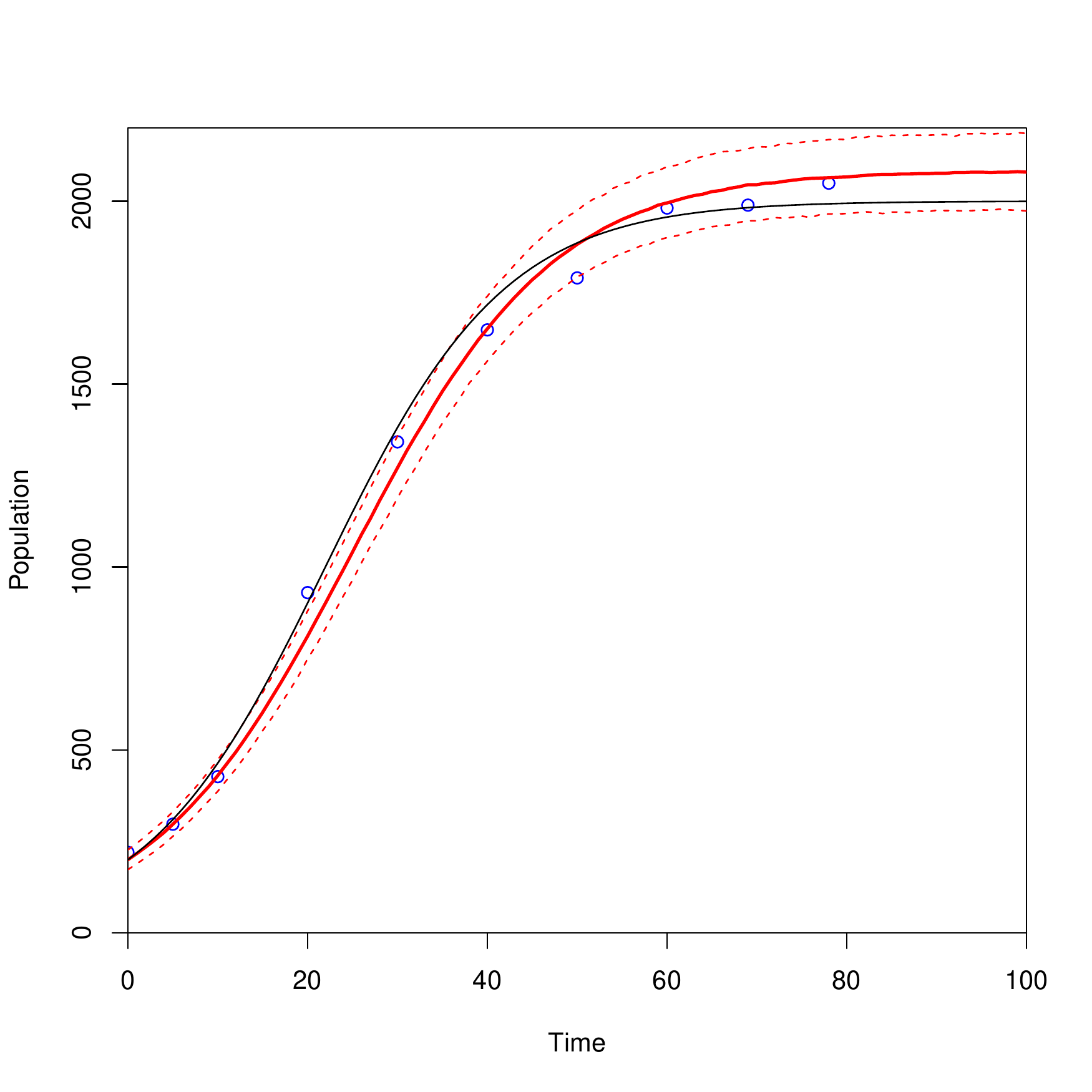}
}
\subfigure[]{
\includegraphics[width=1.96in]{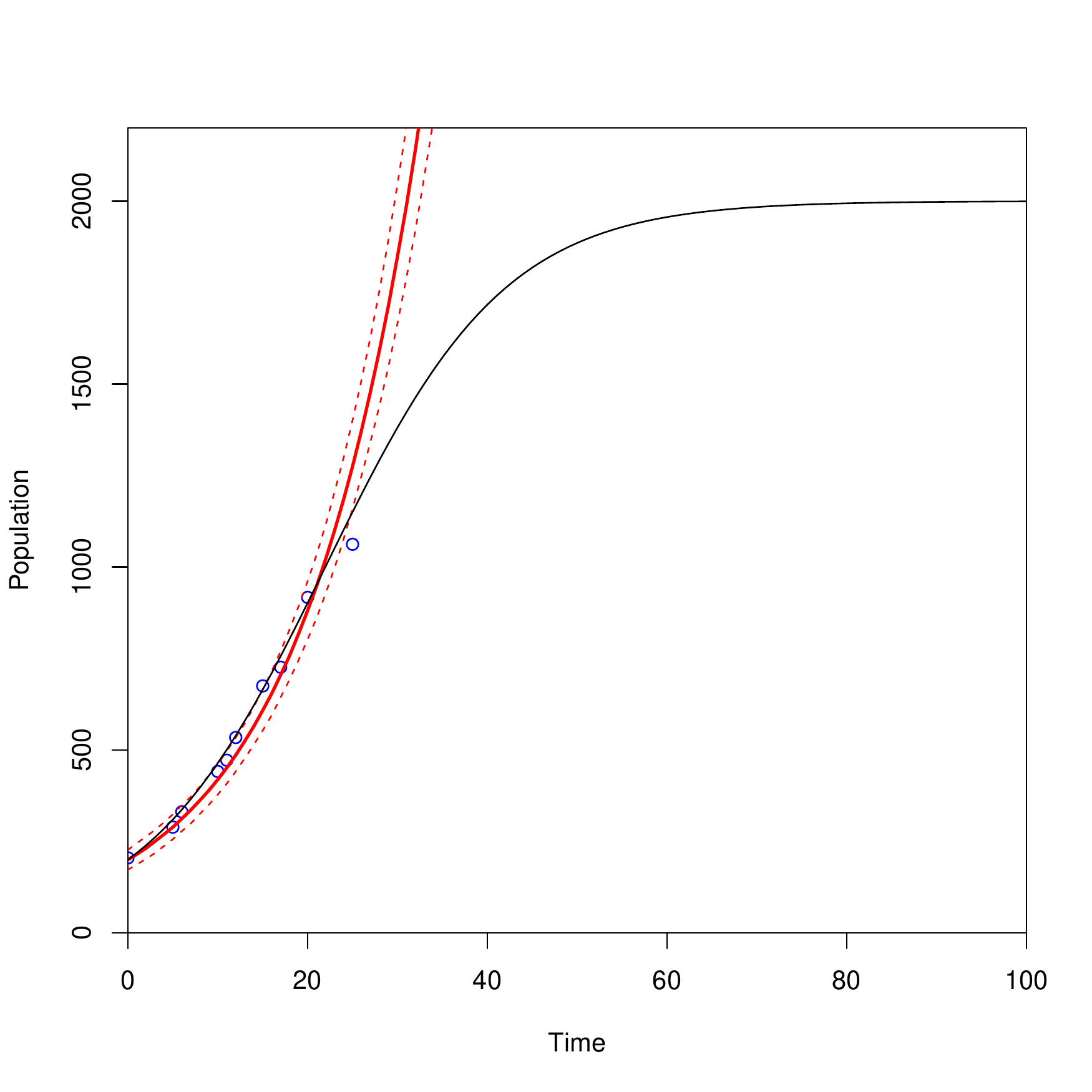}
}
\subfigure[]{
\includegraphics[width=1.96in]{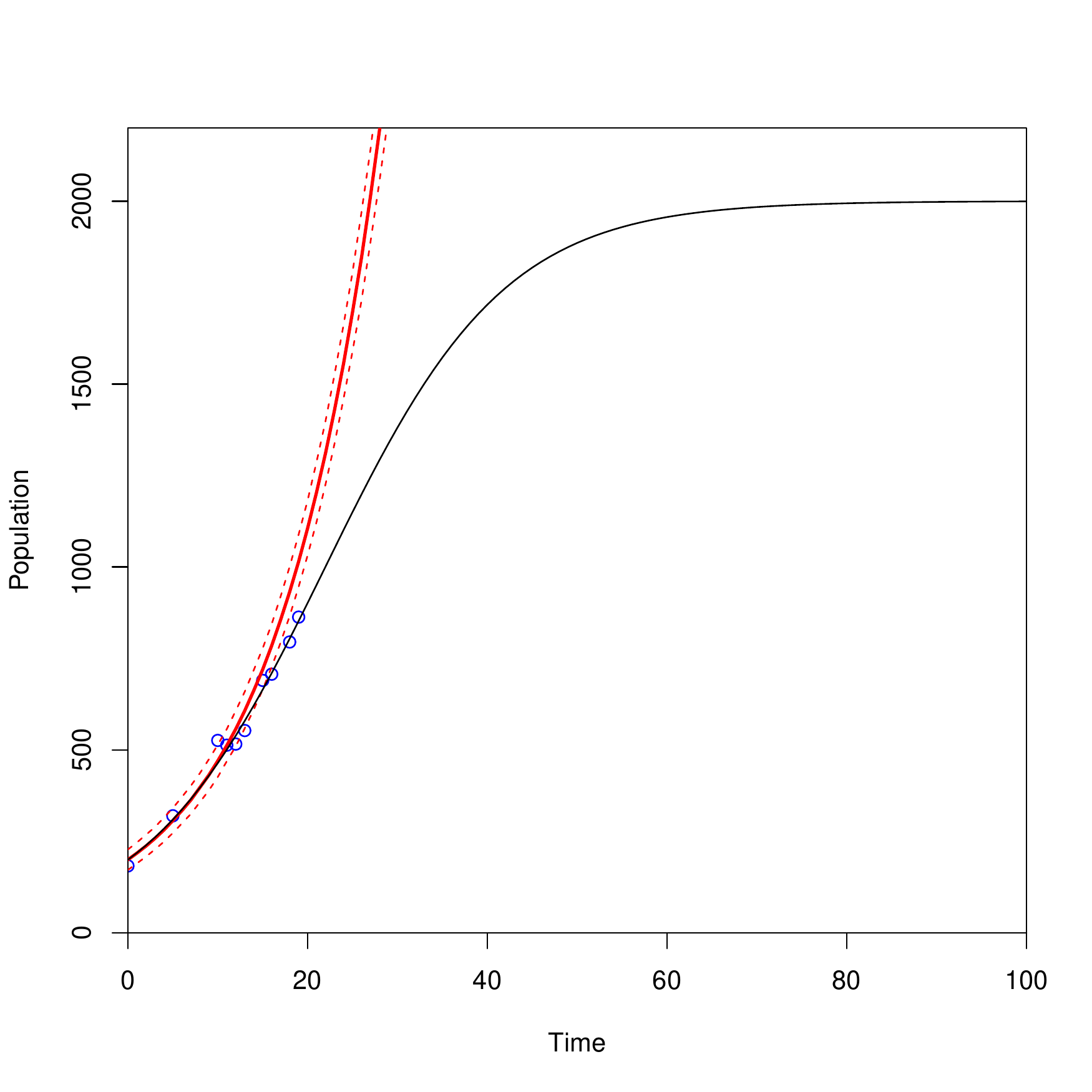}
}
\caption{10,000 MCMC samples simulate a normal growth curve increasing at a 10\% rate across 100 time points with a carrying capacity of 2000. Sequential optimality is used with the Bayesian design criteria to find the ten best points for (a) $\bar{U_I}$, (b) $\bar{U_A}$ and (c) $\bar{U_D}$ optimal designs. The optimal points are plotted in blue and fit with a solid red curve. The red dotted lines represent the 2.5\% and 97.5\% quantiles of the predicted values of the model. The black curve represents the ground truth model.}
\label{fig:SQ6}
\end{figure}

The frequency plots in Figure \ref{fig:SQ5} show the probability densities associated with the first set of candidate points evaluated in the sequential optimality algorithm. The Bayesian I-optimal design gives weight to specific candidate points, whereas the Bayesian A and D optimality criteria provide a wide range of optimal candidates. Based on the frequency charts, it is clear that the Bayesian I-optimal design can provide a precise optimal design point. Whereas, the Bayesian A and D optimal designs appear to lack precision. These results are further visualized by plotting the Bayesian sequential designs in Figure \ref{fig:SQ6} for the normal growth model. 

Given all of the criteria and varying growth rates, the simulations using sequential optimality resulted in different designs. Some designs are able to capture the dynamics of the system while others require a larger design point budget. Ultimately, this indicates that there are several ways to design experiments for dynamic models. Using the simple example of population growth modeled by the logistic equation, the sequential optimality simulations demonstrate an adaptation of the commonly performed simulated annealing algorithm. Thus, sequential optimality can be recommended when limitations arise for sampling.

\section{Discussion}

In this study, the dynamics of various population growth models were investigated in order to design optimal sampling regimes for ecologists. Efforts were focused on a simple model with limited external factors to demonstrate a new and novel approach of sequential optimality. For the purpose of this study, the logistic equation provided a straightforward model used to compare sampling techniques. Simulated annealing was implemented to explore the design space using various optimality criteria. Despite the growth rate and criterion used, simulated annealing was able to capture the dynamics of the models by exploring the entire design region. However, sequential optimality found sampling regimes that capture the dynamics of a system in a sequential manner. 

The proposed method of sequential optimality incorporates prior information into the design process. Rather than exploring the entire design space at once, the sampling method evaluates subsets of the region using prediction based criteria. Sequential optimality captured the dynamics of a normal growth rate using the I optimality criterion. However, implementing this method across criteria and growth rates led to different results. The normal growth rate model was captured by I and $A_{\Phi}$ optimal designs but required more design points for the $D_{\Phi}$ optimal design. Models with fast growth rates were always captured leading to the belief that a smaller design point budget could be explored. On the other hand, slow growth rates required a larger design point budget no matter the criteria used. The algorithm had a more difficult time distinguishing the dynamics of the slower growth rate.

Bayesian optimal designs were also considered as a method for comparison. The $\bar{U_I}$-optimality criterion had more precise candidate points than the $\bar{U_A}$ or $\bar{U_D}$ optimality criteria. This led to the results that only the $\bar{U_I}$ Bayesian optimal design was able to capture the dynamics of the curve. The other two designs required more design points to capture the dynamics of the model. Comparing Bayesian sequential designs to the traditional criteria provided yet another technique available for determining the optimal design points in the region. Comparing these processes across criteria and models suggests that there are several ways to design experiments. However, the approach of sequential optimality is beneficial given that the procedure predicts the next optimal design point, which could be helpful when sampling budgets change. 

\section{Conclusions}

In this paper, the ecological model known as the logistic equation was used based on the various applications and popularity of the model. The proposed design space tracked the change in population dynamics across time and was explored using our novel approach of sequential optimality. When using prediction based criteria for a normal growth model, we captured the dynamics of the system and found an optimal design that translates to an optimal sampling regime for ecologists. Rather than sampling out of convenience or across an equal interval, we were able to demonstrate a method that can provide the next optimal time to sample given the current state of the environment. Real data cannot be incorporated at this time given that this paper introduces a new method for designing sampling regimes. However, the developed method does learn about processes in a sequential manner and has the potential to assist ecologists when planning sampling schedules. 

Though the analysis is performed on a straightforward univariate model, we acknowledge that more complex dynamics exist and can represent more realistic environmental encounters. Incorporating external factors into the model could lead to substantial improvements to our method. Furthermore, there are many design problems that can be investigated by a Bayesian approach. Specifically related to sequential optimality, this study focused on designs of a set size that explore a set window of time into the future. The method could be expanded upon by studying varying design sizes and design windows. Also, sampling techniques used by ecologists when fishing, farming, or testing germination could be invasive and change the process. These changes could be taken into consideration as they may affect the model. In this paper, we focused on providing optimal designs based on temporal models. However, in the future we could explore spatial models or create a hierarchical network of temporal models. Given these additional conditions, there are many levels of uncertainty that we could incorporate to reflect the reality of an ecological system. 


\bibliographystyle{spbasic}      
\bibliography{bib1}   

\end{document}